\documentclass[aps,prb,reprint,superscriptaddress,showpacs]{revtex4-1}
\usepackage{graphicx}

\begin{document}

\title{Magnetic interactions in disordered perovskite PbFe$_{1/2}$Nb$_{1/2}$O$_{3}$
and related compounds. Dominance of nearest-neighbor interaction.}

\author{R.O.\ Kuzian}

\affiliation{Institute for Problems of Materials Science NASU, Krzhizhanovskogo
3, 03180 Kiev, Ukraine}

\affiliation{Donostia International Physics Center (DIPC), ES-20018 Donostia-San
Sebastian, Spain}

\author{I.V.\ Kondakova}

\affiliation{Institute for Problems of Materials Science NASU, Krzhizhanovskogo
3, 03180 Kiev, Ukraine}

\author{A. M. Dar\'e}

\affiliation{Aix-Marseille Universit\'e, CNRS, IM2NP UMR 7334, F-13397, 
Marseille, France
}

\author{V.V.\ Laguta}

\affiliation{Institute for Problems of Materials Science NASU, Krzhizhanovskogo
3, 03180 Kiev, Ukraine}

\affiliation{Institute of Physics, AS CR, Cukrovarnicka 10, 16253 Prague, Czech
Republic}

\date{26.11.13}

\begin{abstract}
We show that the magnetism of double perovskite 
AFe$_{1/2}$M$_{1/2}$O$_{3}$ 
systems may be described by the Heisenberg model on the simple
cubic lattice, where only half of sites are occupied by localized
magnetic moments. The nearest-neighbor interaction $J_{1}$ is more
than 20 times the next-nearest neighbor interaction $J_{2}$, the
third-nearest interaction along the space diagonal of the cube being
negligible. We argue that the variety of magnetic properties observed
in different systems is connected with the variety of chemical ordering
in them. We analyze six possible types of the chemical ordering in
2$\times$2$\times$2 supercell, and argue that the probability to find 
them in a real compound does not correspond to a random occupation of 
lattice sites by magnetic ions. The exchange  $J_{2}$ rather than 
$J_{1}$ define the 
magnetic energy scale of most double perovskite compounds that means
the enhanced probability of 1:1 short range ordering.
Two multiferroic compounds PbFe$_{1/2}$M$_{1/2}$O$_{3}$ (M=Nb, Ta)
are exceptions. We show that the relatively high temperature of 
antiferromagnetic transition is compatible with a
layered short-range chemical order, which was recently shown to be most 
stable for these two compounds 
[I. P. Raevski, {\em et al.}, Phys.\ Rev.\ B \textbf{85}, 224412 (2012)].
We show also that one of the types of ordering has ferrimagnetic
ground state. The clusters with short-range order of this type may
be responsible for a room-temperature superparamagnetism, and may
form the cluster glass at low temperatures. 
\end{abstract}


\maketitle

\section{Introduction.}

\begin{figure}
\includegraphics[width=0.85\columnwidth]{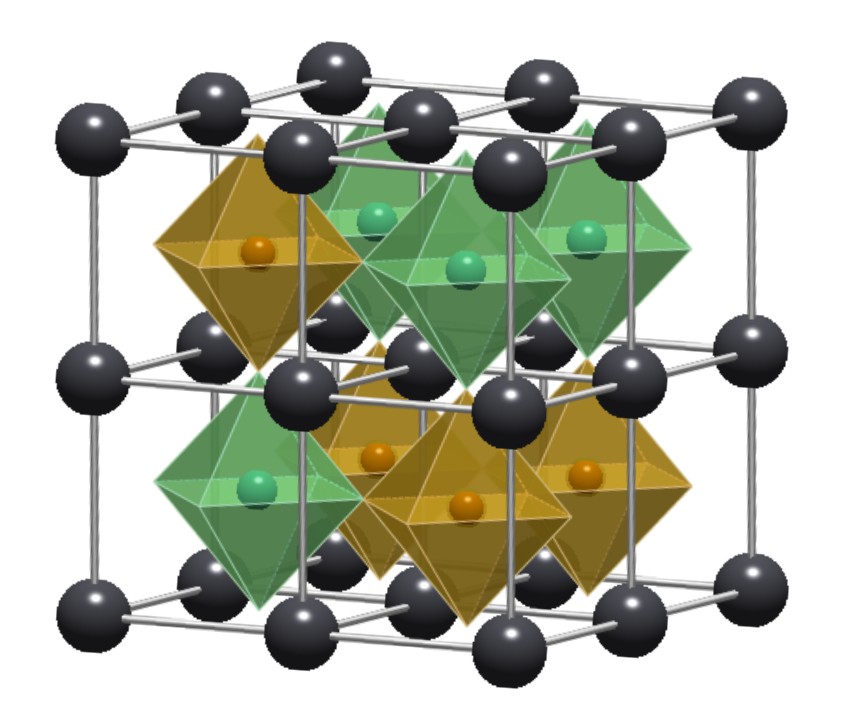} 
\caption{(Color online)
A fragment of disordered PbFe$_{1/2}$Nb$_{1/2}$O$_{3}$ structure. 
The supercell containing $8 = 2\times 2\times 2$  perovskite cells
is shown. Black circles denote Pb ions, 
green (brown) circles inside oxygen octahedra depict Fe (Nb) ions. 
Oxygen ions are located in the corners of the  octahedra.
 The distribution of Fe and Nb ions corresponds to PFB4
chemical order (see text).}
\label{scell}
\end{figure}

The compound PbFe$_{1/2}$Nb$_{1/2}$O$_{3}$ (PFN)
is one of the first multiferroics reported\cite{Smolenskii58,Bokov62}. It
remains to be in focus of the attention of multiferroic community.
\cite{Raevski09,Rotaru09,Kleemann10,Laguta10,Raevski12,Laguta13,Kubrin}
Despite the long story of studies, the magnetic properties of PFN are not
fully understood. It belongs to the family of
double perovskites AFe$_{1/2}$M$_{1/2}$O$_{3}$=A$_2$FeMO$_6$  with a
nonmagnetic cation in A site (A=Pb,Ca,Sr,Ba) of the perovskite 
structure ABO$_3$
and a distribution of the magnetic Fe$^{3+}$ and a nonmagnetic M$^{5+}$
cations (M=Nb,Ta,Sb) in six-coordinated B-site of the structure (see 
Fig.~\ref{scell}).

The magnetic properties of these compounds are defined by Fe$^{3+}$,
$S=5/2$ ions that occupy half of sites of simple cubic lattice (sublattice
B of perovskite structure), and interact via various superexchange paths.

It is natural to compare the magnetism of AFe$_{1/2}$M$_{1/2}$O$_{3}$ 
compounds with ortoferrites
RFeO$_3$ (R=Y or a rare earth) and bismuth ferrite BiFeO$_3$, with
a similar perovskite structure where Fe occupy every B site. All these
compounds exhibit essentially antiferromagnetic ordering (with a small
canting of predominantly antiferromagnetic spins) below
the transition temperature, which varies in the range 620$<T_N(1)<$740~K.
\cite{Treves65,Eibschuetz67,Gabbasova91}
The nearest-neighbor Fe-Fe interaction (via Fe-O-Fe path) was estimated as
$J_{1}\sim 50$~K,\cite{Eibschuetz67,Gorodetsky69,Shapiro74,Gukasov97,
Delaire12,McQueeney08,orthofe} the next-nearest-
neighbor being much smaller $\alpha = J_2/J_1 \simeq 0.05$.
\cite{Gorodetsky69,Shapiro74,Gukasov97}

If one assume (i) a random occupation of the site B by Fe and M ions 
(the X-ray
diffraction and M\"{o}ssbauer spectra support this assumption for most of
M ions), and (ii) a similar value of Fe-O-Fe superexchange, we may 
expect the
N\'eel temperature $T_N(0.5) \sim 0.5T_N(1) >300$~K. This estimate 
comes from
an analogy with $T_N(x)$ behaviour in the disodered perovskite system
KMn$_x$Mg$_{1-x}$F$_3$,\cite{DAriano82,Breed70} which agree with
theoretical considerations of dilute Heisenberg magnets.
\cite{Stinchcombe79,Kumar81}
Contrary to these expectations, most of
the AFe$_{1/2}$M$_{1/2}$O$_{3}$ 
compounds exhibit a magnetic anomaly at $T\sim 25$~K.
\cite{Battle95,Battle95a} One observe $T_N \sim 150$~K only for
PbFe$_{1/2}$M$_{1/2}$O$_{3}$ (M=Nb,Ta).\cite{Smolenskii58}
It seems that at least one of the above assumptions (i), (ii) is false.

\begin{figure}
\includegraphics[width=0.32\columnwidth]{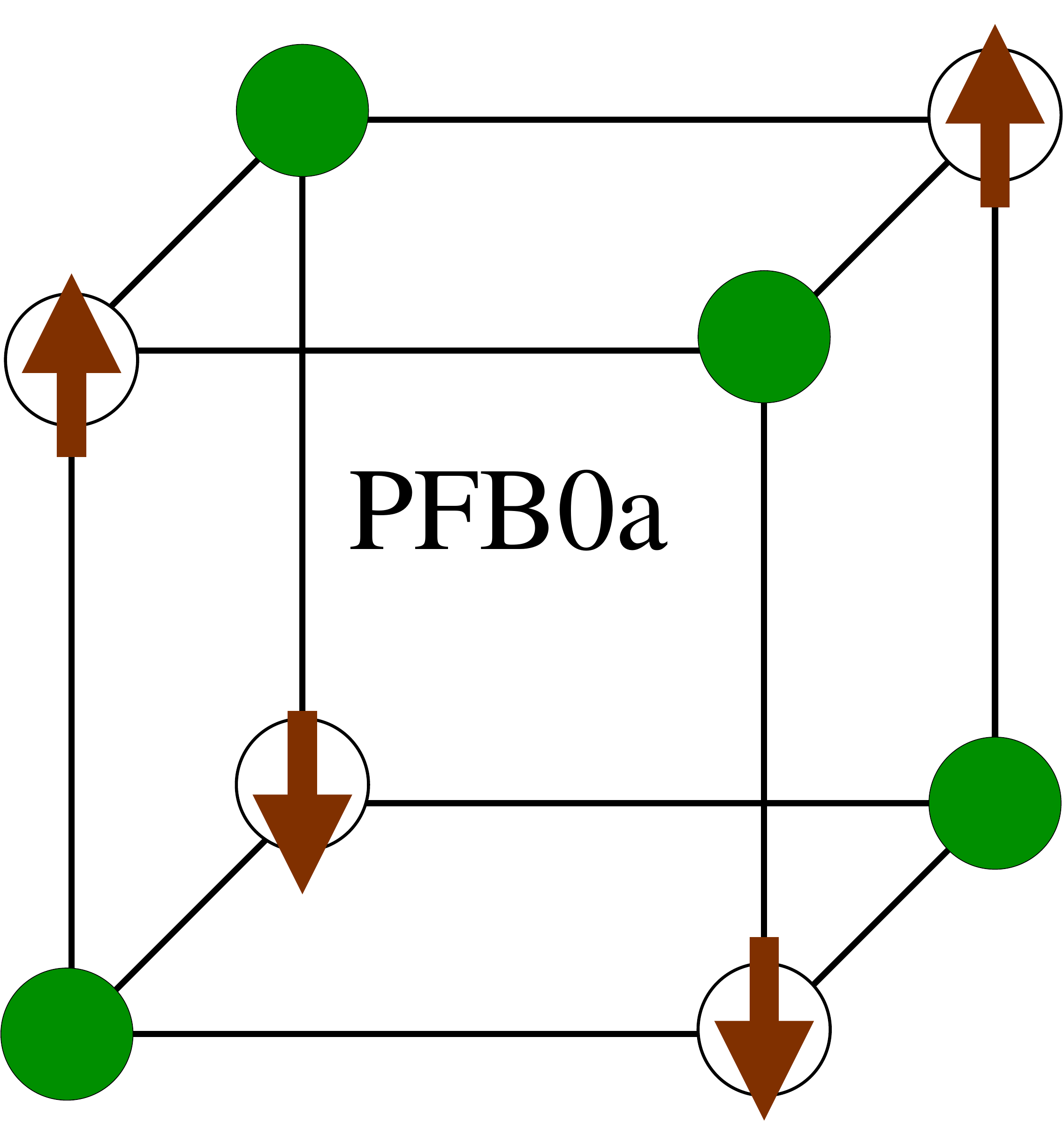} 
\includegraphics[width=0.32\columnwidth]{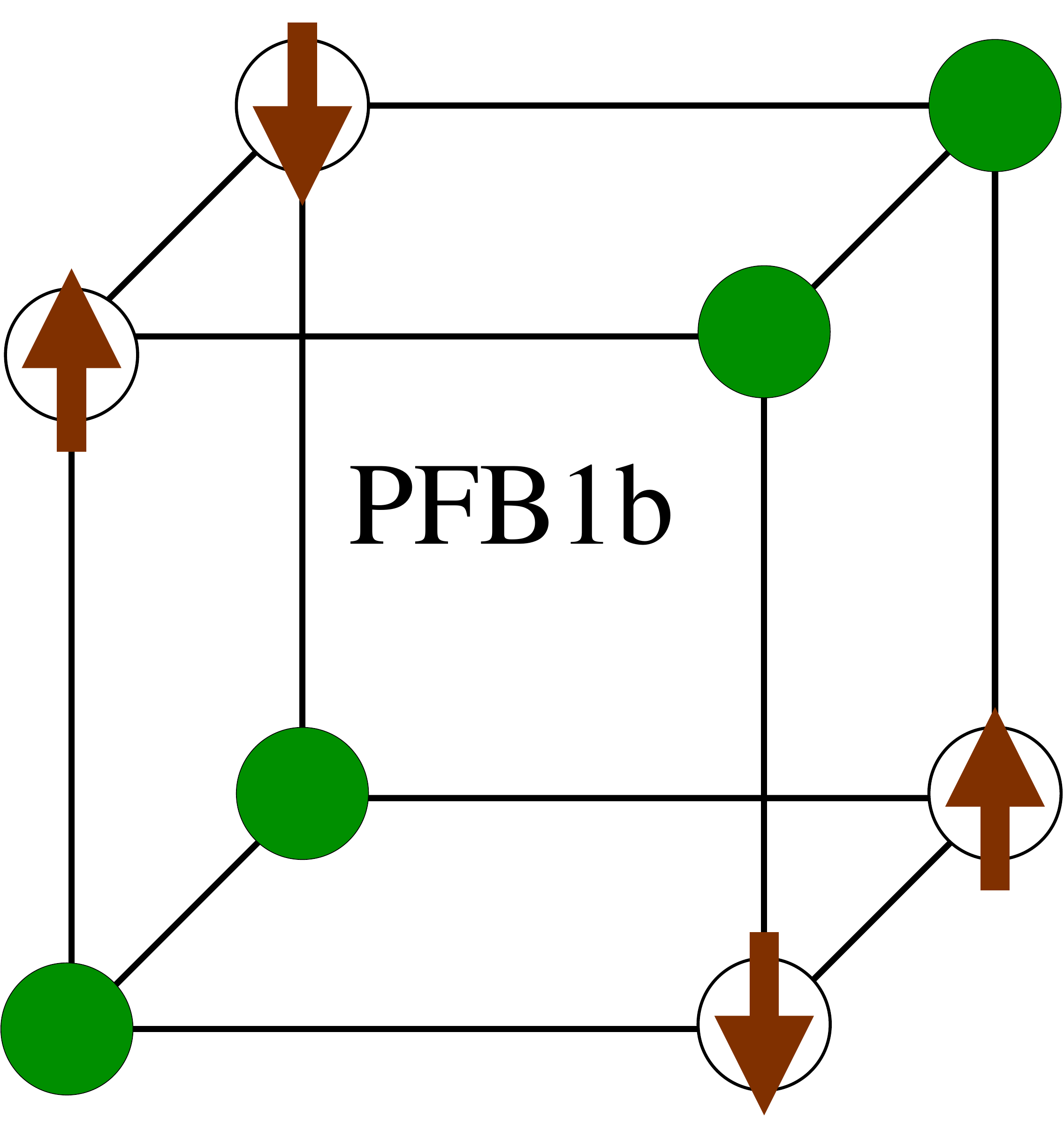}
 \includegraphics[width=0.32\columnwidth]{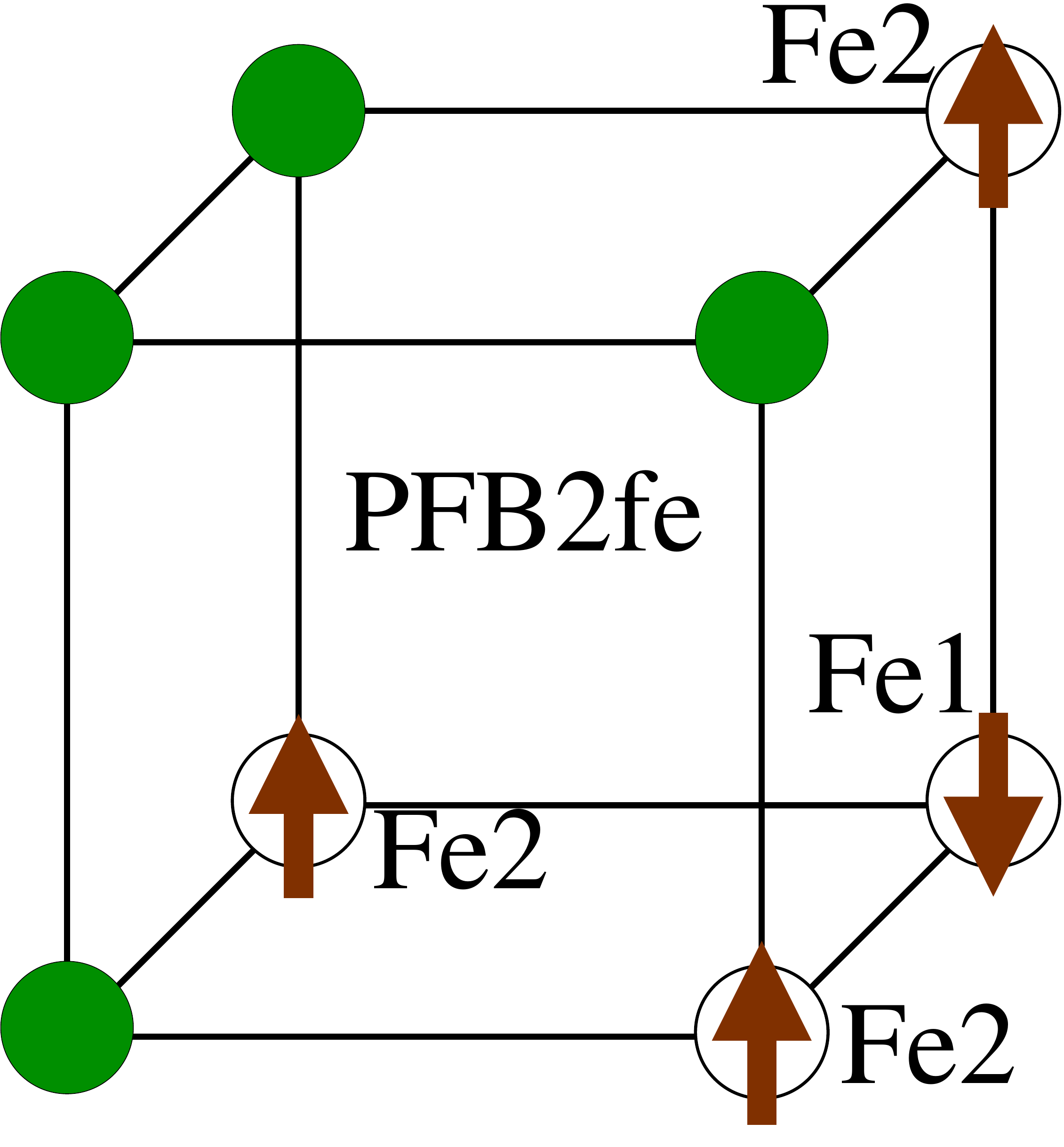}\\ 
 \includegraphics[width=0.32\columnwidth]{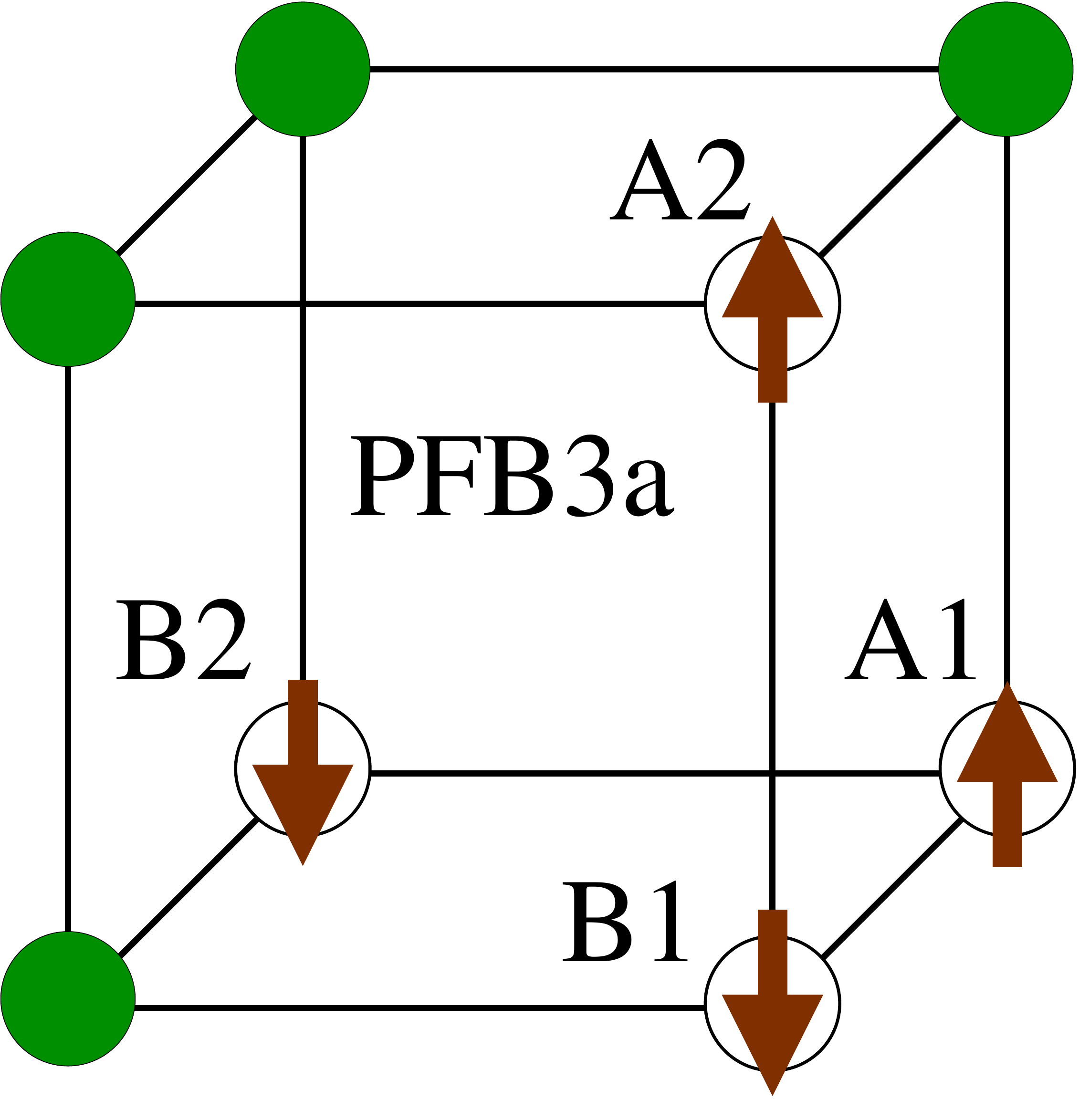}
 \includegraphics[width=0.32\columnwidth]{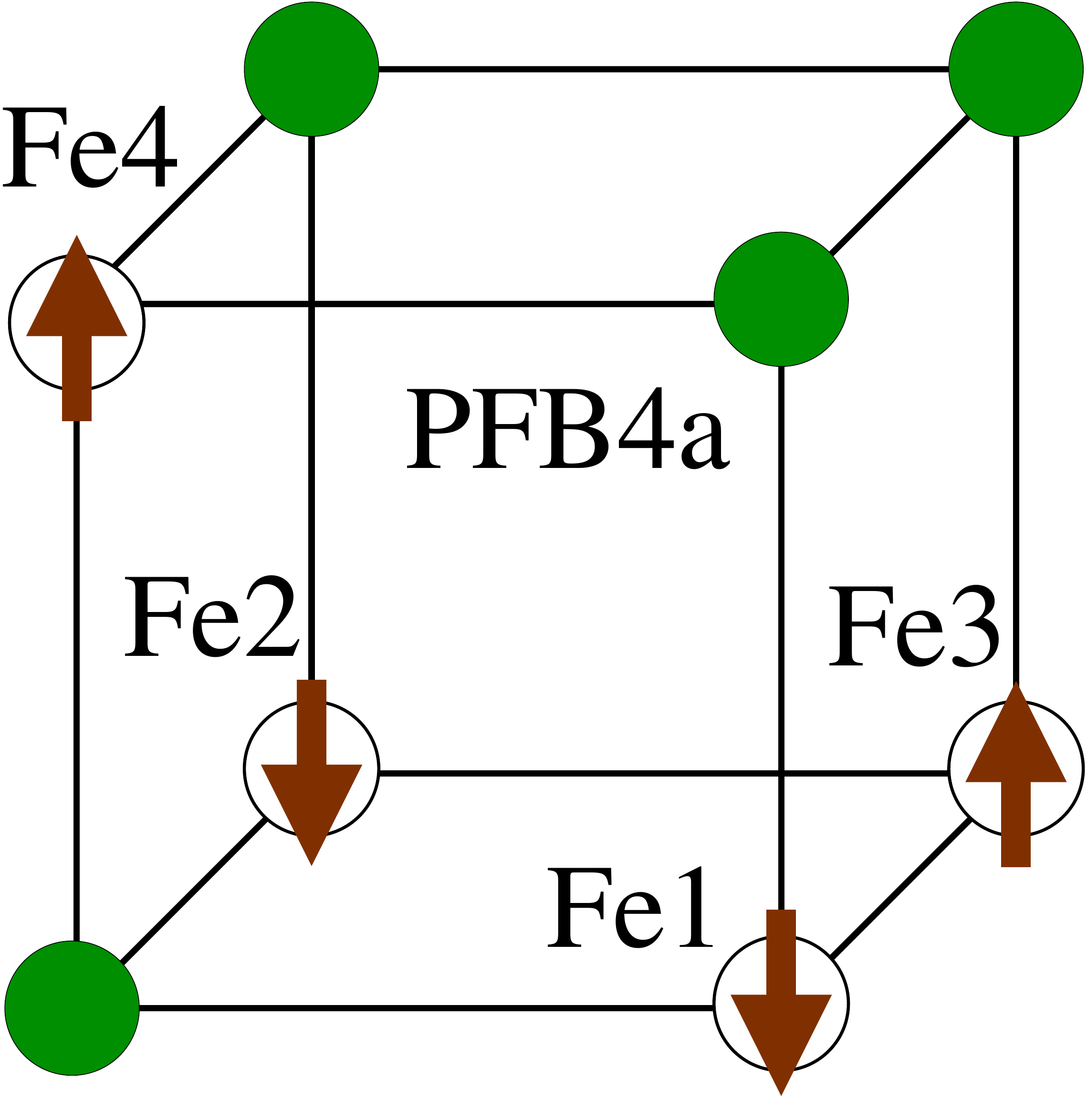} 
 \includegraphics[width=0.32\columnwidth]{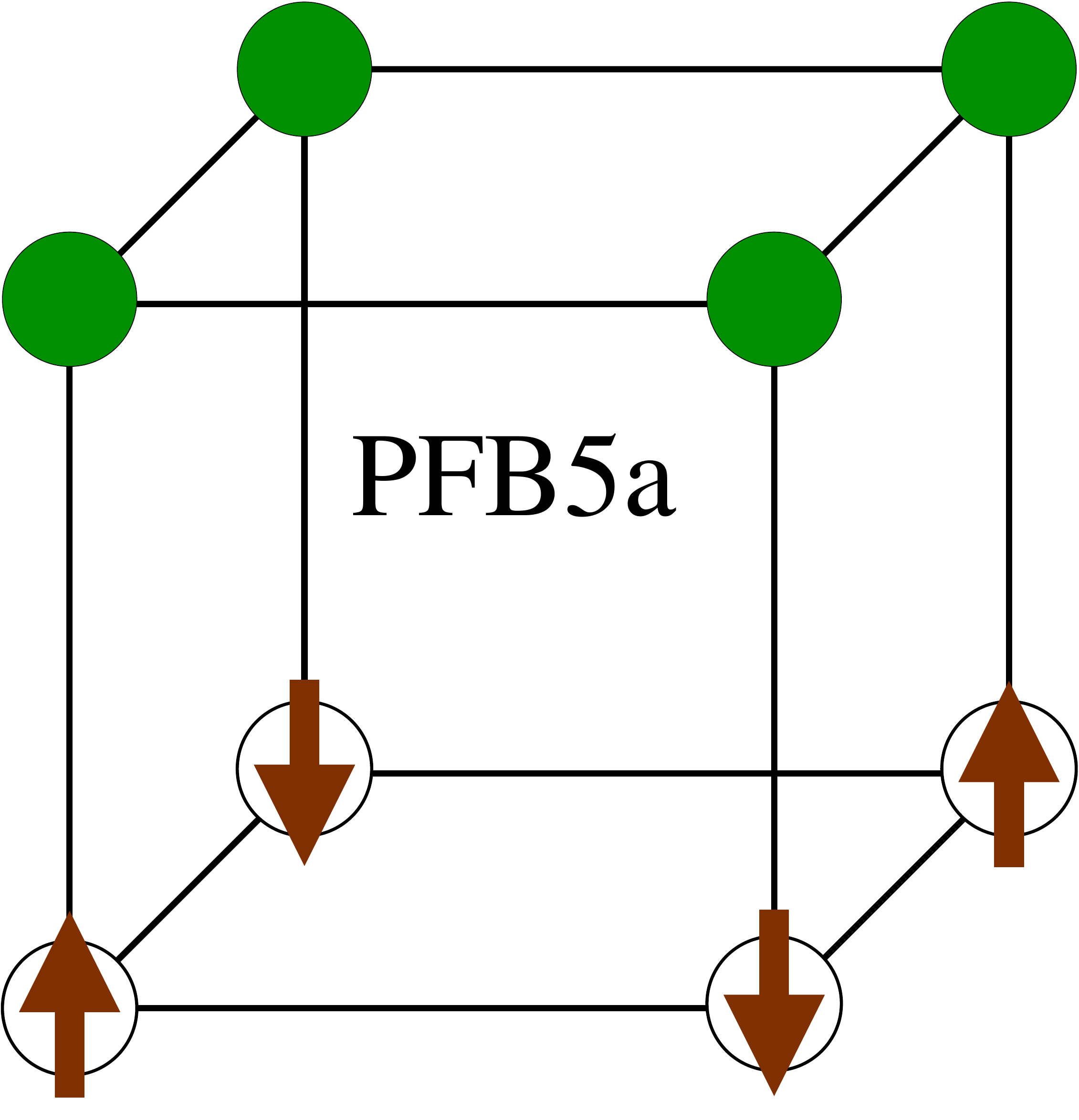}
 \caption{(Color online)
 Magnetic ground states for different chemical configurations 
 of Fe$^{3+}$ (open circles) in
a 2$\times$2$\times$2 supercell of AFe$_{1/2}$Nb$_{1/2}$O$_{3}$ (A=Pb,Ba,
only B-sublattice sites are shown). Green filled circles denote 
non-magnetic Nb$^{5+}$ ions.
PFB0a is the I-type order of the fcc lattice.}
\label{mstr}
\end{figure}

Evidences for partial chemical ordering in B sublattice comes from
experiment\cite{Battle95,Battle95a,Blinc08,Laguta10} and 
theory.\cite{Gu99,Raevski12} 
The disorder in the distribution of Fe and M ions was modeled
in the Ref.~\onlinecite{Raevski12,rtab} 
by a set of 6 periodic lattices PFB0\dots PFB5 
with the supercell containing
$8 = 2\times 2\times 2$  perovskite cells with different versions
of chemical order (ion distributions) within the cells 
(see Fig.~\ref{mstr}).
It was shown that the total energy is
different for different configurations \cite{rtab}, and
the hierarchy of the energies depends on the type of M-ion.

Recent reports on room-temperature multiferroicity of 
PFN/PbZr$_x$Ti$_{1-x}$O$_{3}$\cite{Sanchez13}, and of 
related solid solution systems
PbFe$_{1/2}$Ta$_{1/2}$O$_{3}$/PbZr$_x$Ti$_{1-x}$O$_{3}$\cite{Sanchez11,Evans13}
and Pb(Fe$_{2/3}$W$_{1/3}$)O$_3$/PbZr$_x$Ti$_{1-x}$O$_{3}$ \cite{Kumar09}
evidences in favor of presence in these systems of magnetic 
interactions $J$ with the energy scale 
$S(S+1)J/k_B=8.75J/k_B \sim 300$~K. In the Ref.~\onlinecite{Lampis04},
the nearest-, second-, and fourth-nearest-neighbor exchange
interactions between Fe$^{3+}$ ions were
found from LSDA$+U$ calculations for PbFe$_{1/2}$Ta$_{1/2}$O$_{3}$. 
The nearest-neighbor exchange 
occurs (in our notations, see Eq.(\ref{eq:Heff})) 
to be $J_1/k_B \approx 42$~K, it gives $S(S+1)J/k_B=366$~K.

In this work, using first-principle calculations,
we find the values of exchange interaction
between nearest-, second-, and third-nearest-neighbor Fe$^{3+}$ ions
in PFN, 
and confirm the validity of the assumption (ii), i.e. we show that the
the nearest-neighbor interaction dominates, and its value is close to
that found for RFeO$_3$ and PbFe$_{1/2}$Ta$_{1/2}$O$_{3}$ compounds. 
So, the peculiarities of magnetic
properties of AFe$_{1/2}$M$_{1/2}$O$_{3}$ 
compounds are related with chemical ordering in
B-sublattice.

\section{Method}

The density functional theory calculations were performed using the
full- potential local-orbital (FPLO) code.\cite{FPLO} 
We have used the default FPLO basis, which is claimed to be technically 
complete, i.e. the FPLO code developers have checked the convergence 
of the electronic density with respect to the 
number of basis functions for a huge number of compounds, including
$3d$-metal oxides.
The FPLO basis consists from localized atomic-like functions 
defined by angular $nl$-quantum numbers and the number of 
numerical radial functions per
orbital. Each valence states can come as single, double or 
tripple state, which means
that there are 1, 2 or 3 radial basis functions for this $nl$-quantum number.
The default basis for Fe is: single 3s3p4p, and double 4s3d,
for O: single 1s3d, double 2s2p, for Pb: single 5s5p5d6d, double 6s6p,
for Nb: single 4s4p5p, and double 5s4d.
The exchange
and correlation potential of Perdew and Wang\cite{PW} was employed
as well as the FPLO implementation of the LSDA+$U$ method in the
atomic limit scheme\cite{Eschrig03,Czyzyk94}, and parameters
$U\equiv F^0=$4 and 6.
The intra-atomic exchange parameters were fixed at the values
$F^2=49B+7C=10.3$~eV, and $F^4=63C/5=7.5$~eV, which corresponds to
Racah parameters $B=1015$~cm$^{-1}$, $C=4800$~cm$^{-1}$ for free Fe$^{3+}$ ion.
\cite{Tanabe54}

\begin{figure}
\includegraphics[width=0.32\columnwidth]{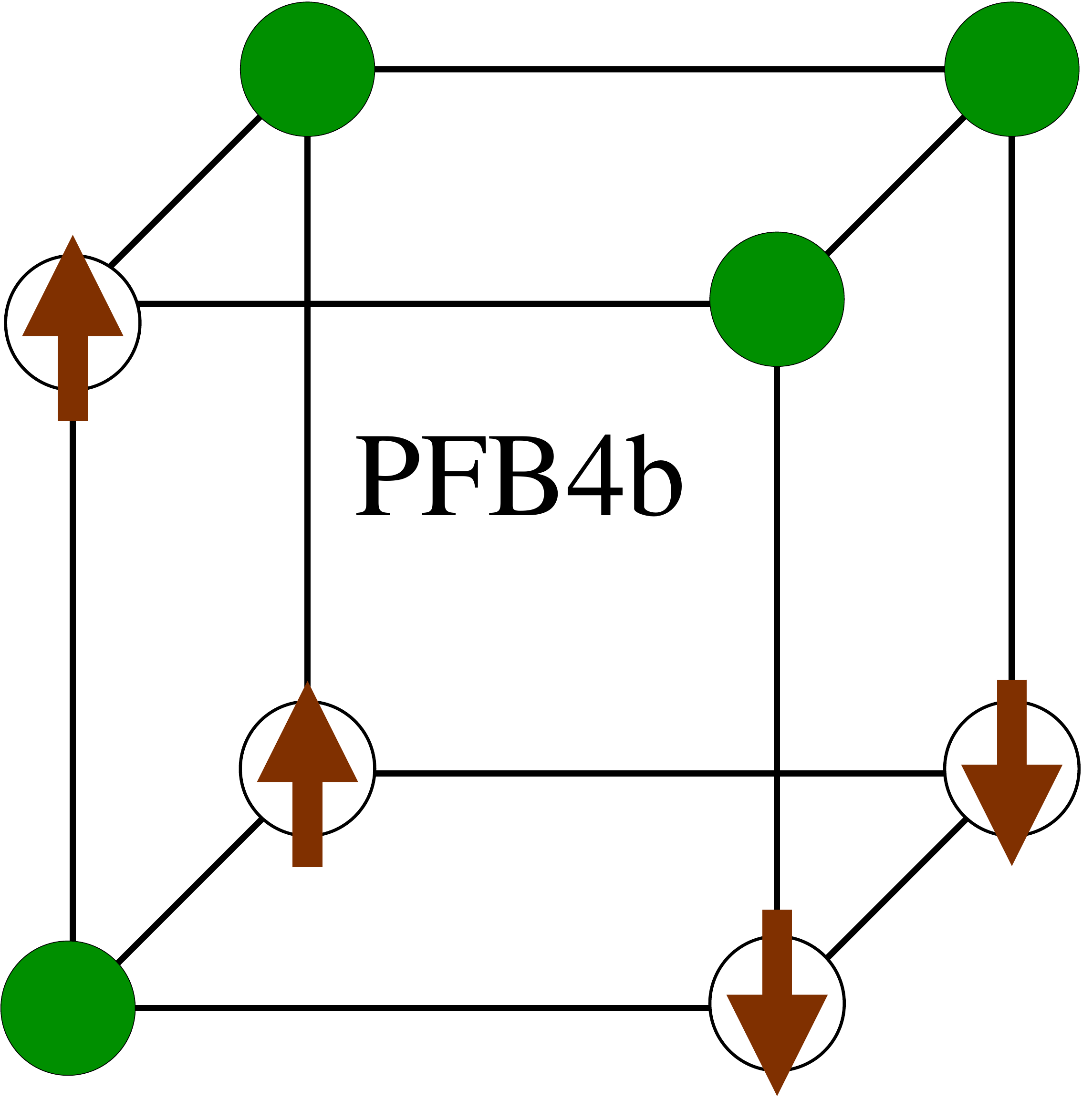} 
\includegraphics[width=0.32\columnwidth]{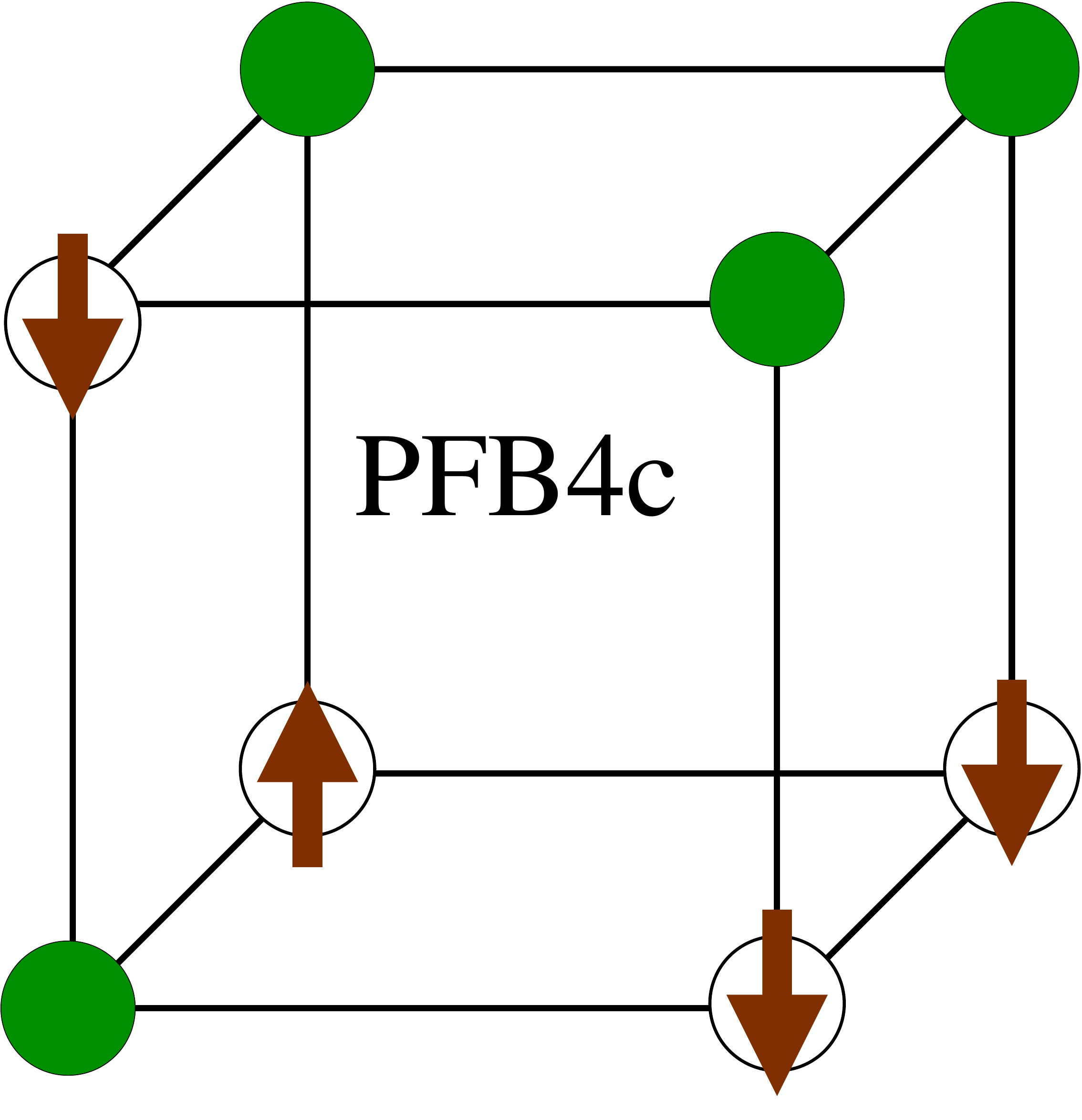} 
\includegraphics[width=0.32\columnwidth]{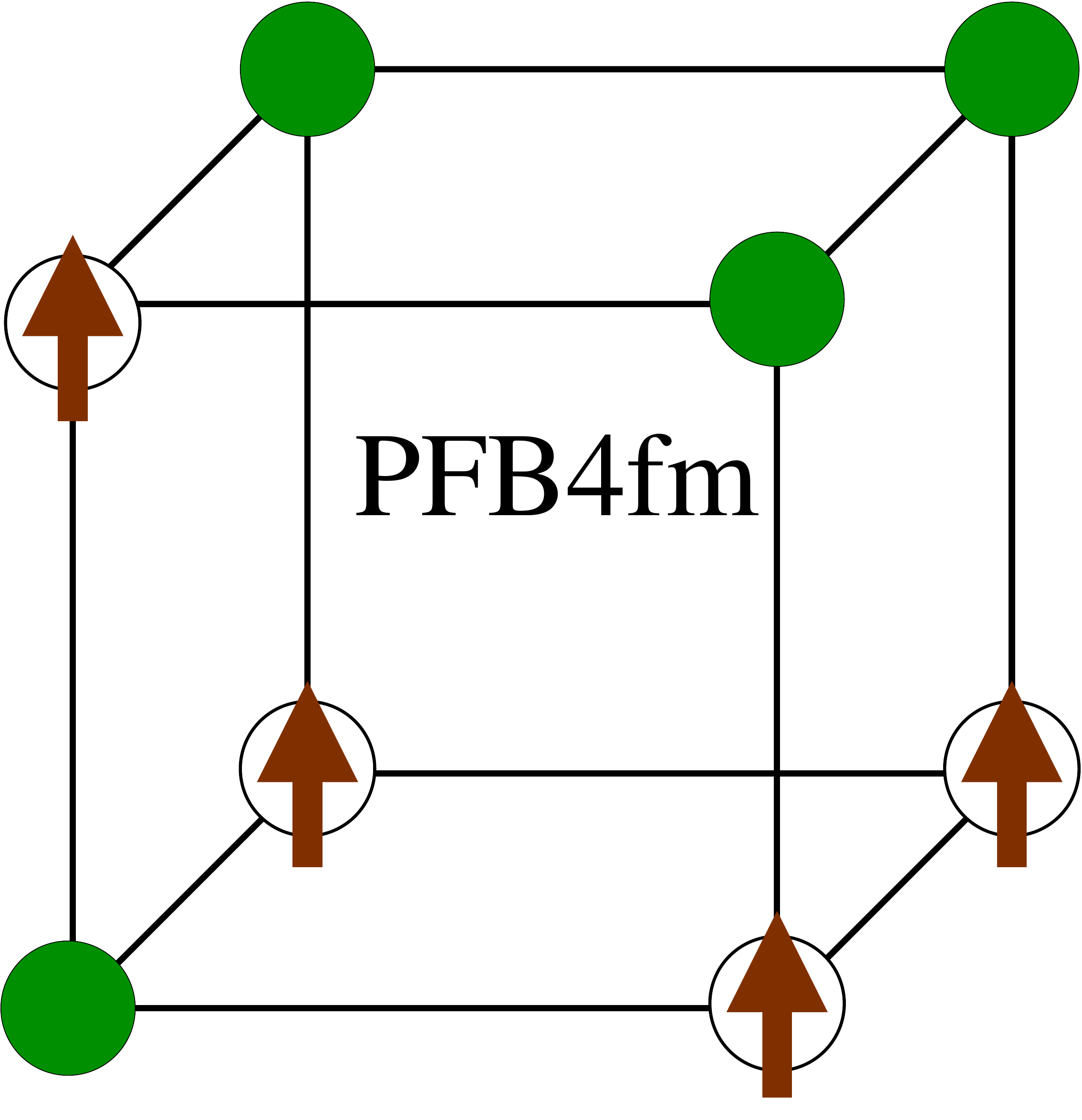}\\
\includegraphics[width=0.32\columnwidth]{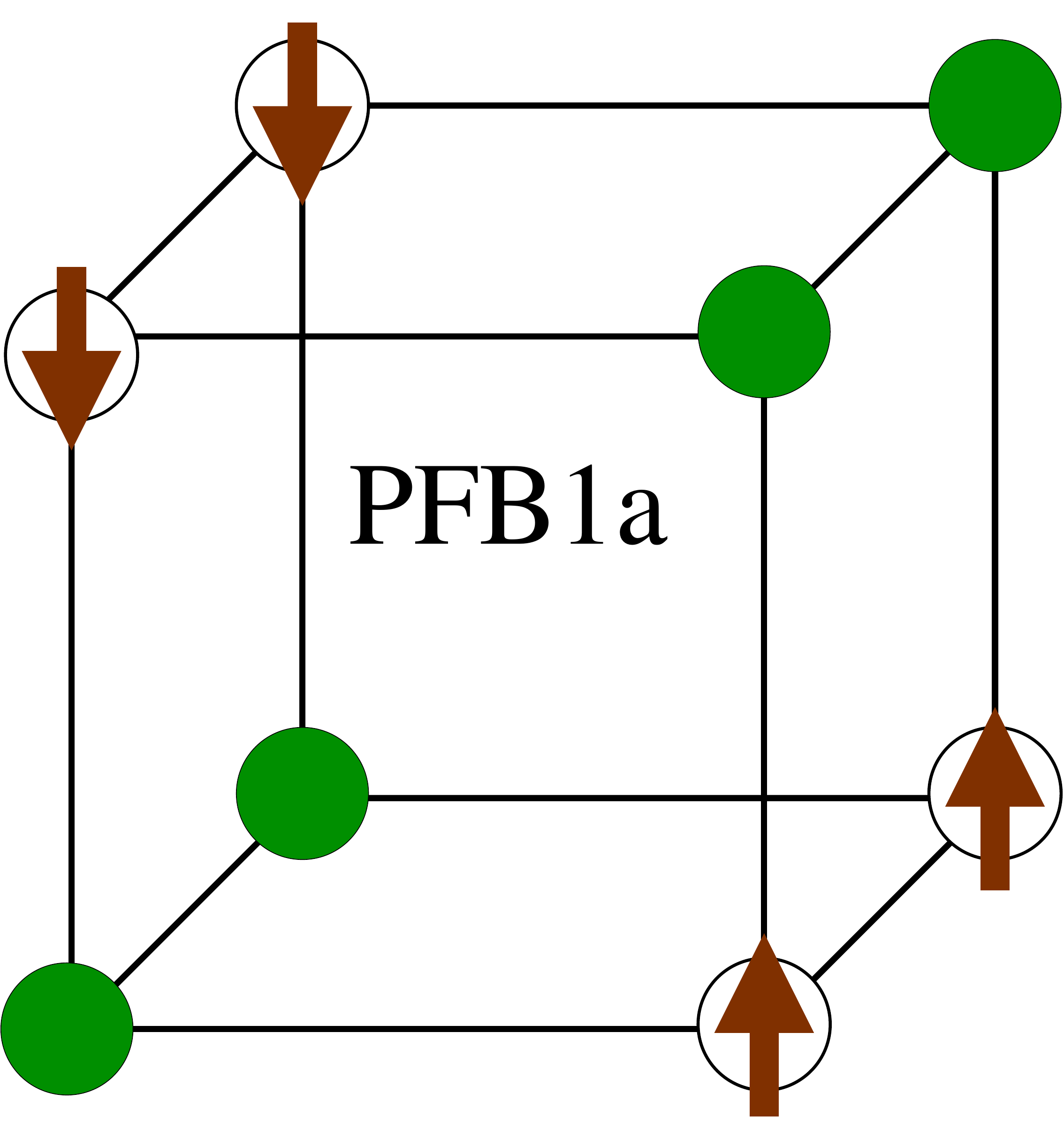}
\includegraphics[width=0.32\columnwidth]{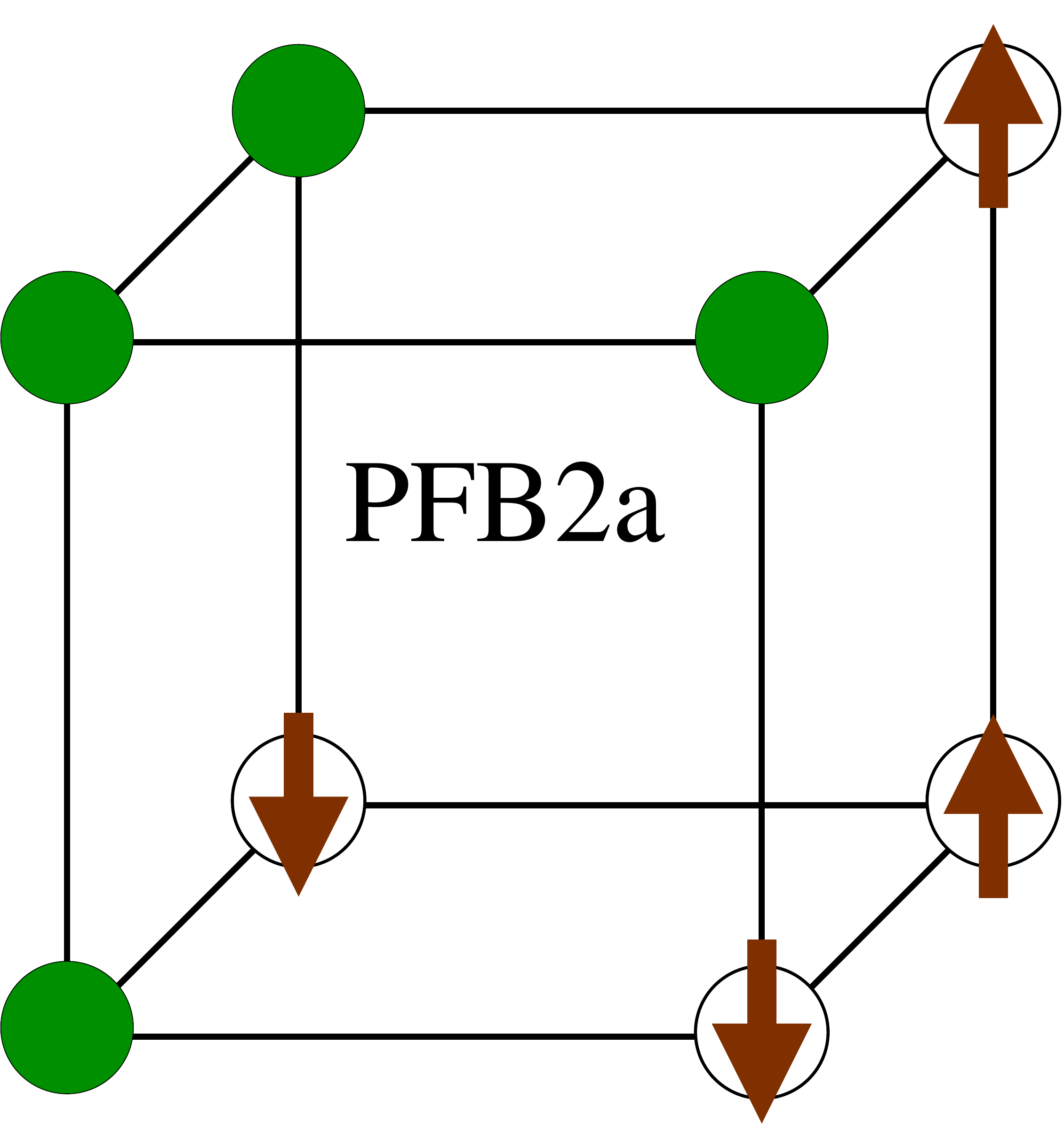}
\caption{(Color online)
Upper panel: The excited magnetic states of PFB4 chemical order that were used for 
the calculations of the interactions.
Lower panel:
Additional magnetic structures, which were used for the check of the 
mapping of LSDA+$U$ on the Heisenberg model (\ref{eq:Heff}).}
\label{mstr2}
\end{figure}

The calculations were made for $2\times2\times2$ 40 atom 
supercell Pb$_{8}$Fe$_{4}$Nb$_{4}$O$_{24}$
(symmetry group P1, \#1) shown shematically 
in the Fig.~\ref{scell}. The $4\times4\times4$ k-mesh was used for
the Brillouin zone integration.
First, we have defined the magnetic interaction
for cubic perovskite structure that corresponds to the paraelectric phase 
of PFN with 
the experimental lattice parameter $a=4.01$~\AA , and PFB4 chemical 
order (Fig.~\ref{mstr2}). 
Then we have checked that the interaction values are essentially the 
same for all kinds of chemical orders and for actual distorted perovskite
structure of PFN.
The ion coordinates for all possible types
of the chemical ordering shown in the Fig.~\ref{mstr} were taken from
the results of full relaxation\cite{raev2} 
that has been performed in the Ref.~\onlinecite{Raevski12}.

The total energies for different structural and magnetic configurations
were obtained, and the results were mapped onto an effective Hamiltonian
\begin{equation}
\hat{H}=E_{n}+\frac{1}{2}\sum_{\mathbf{R,g}}J_{\mathbf{g}}
\hat{\mathbf{S}}_{\mathbf{R}}\hat{\mathbf{S}}_{\mathbf{R+g}},
\label{eq:Heff}
\end{equation}
 where $E_{n}$ is a non-magnetic, spin-independent part of the energy,
which depends on chemical configuration.\cite{Raevski12} The spin-dependent
part of the interaction has the form of a Heisenberg term. The sum
goes over the lattice sites $\mathbf{R}$ occupied by magnetic Fe$^{3+}$
ions, vectors $\mathbf{g}$ join interacting spins. The $2\times2\times2$
supercell allows to determine the values of nearest-, second-nearest-,
and third-nearest- neighbor interactions $J_{1},J_{2},J_{3}$, which
corresponds to sites separated by the edge, face diagonal, and space
diagonal of perovskite unit cell. For a given spin configuration,
the total energy per supercell is
\begin{equation}
E_{c}=\left\langle \hat{H}\right\rangle =
E_{n}+\frac{1}{2}\sum_{\mathbf{s,g}}J_{\mathbf{g}}\left\langle
\hat{\mathbf{S}}_{\mathbf{s}}\hat{\mathbf{S}}_{\mathbf{s+g}}
\right\rangle ,
\label{eq:Ecell}\end{equation}
 where $\mathbf{s}$ is the magnetic ion position within the cell,
$\left\langle \hat{\mathbf{S}}_{\mathbf{s}}
\hat{\mathbf{S}}_{\mathbf{s+g}}\right\rangle =cS^{2}$,
$c=+1(-1)$ for parallel (anti-parallel) spin arrangement.

\section{Results}

The details concerning the calculated electronic structure of PFN are given in
the Appendix~\ref{AppA}. Here we concentrate on the magnetic interations.
The results for the total energy calcualtions for different spin
arrangement in  PFB4 chemical order (Figs.~\ref{scell}-\ref{mstr2})
of ideal cubic perovskite structure
are given in the Table~\ref{tab:PFB4}.
\begin{table}
\caption{\label{tab:PFB4}
Total energy differences $E(a,U)$ (meV) for various spin structures, lattice
parameters $a$, and Coulomb repulsion values $U$. The
LSDA+$U$ calculations were performed for ideal cubic 
perovskite structure and  
PFB4 chemical order (see Figs.~\ref{scell}-\ref{mstr2}). 
}
\begin{ruledtabular} \begin{tabular}{llll}
spin str. & $E(4.01,4)$ & $E(3.95,4)$ & $E(4.01,6)$ \\
\hline
PFB4a     & 0           &  0          & 0 \\
PFB4b     & 200         & 213         & 163 \\
PFB4c     & 199         & 211         & 162 \\
PFB4,FM   & 435         & 464         & 341
\end{tabular}\end{ruledtabular}
\end{table}
The expressions for the magnetic energy for the considered supercells
are given in the second column of the Table~\ref{tab:I}. The third
column of the table gives the energies that we obtained in LSDA+$U$
calculations for fully relaxed supercells\cite{Raevski12,raev2}.
\begin{table}
\caption{\label{tab:I} The energies of chemical and magnetic configurations
for Pb$_{8}$Fe$_{4}$Nb$_{4}$O$_{24}$ supercell, which allow to
find all exchange interactions. The calculated LSDA+$U$-values are given for
$U=4$~eV for ion coordinates from Refs.~\onlinecite{Raevski12,raev2}
}
\begin{ruledtabular} \begin{tabular}{lrlr}
conf.  & & $\left\langle \hat{H}\right\rangle $  & $E_{calc}$, meV\\
\hline
PFB0,FM  & $E_{0,fm}=$ & $24J_2S^{2}$  & 69\tabularnewline
PFB0a  & $E_{0,a}=$ & $-8J_{2}S^{2}$  & 0\tabularnewline
PFB1, FM  & $E_{1,fm}=$ & $\left(4J_{1}+8J_{2}+16J_{3}\right)S^{2}$  & 936\\
PFB1a  & $E_{1,a}=$ & $\left(4J_{1}-8J_{2}-16J_{3}\right)S^{2}$  & 903\\
PFB1b  & $E_{1,b}=$ & $\left(-4J_{1}-8J_{2}+16J_{3}\right)S^{2}$  & 437\\
PFB2, FM  & $E_{2,fm}=$ & $\left(6J_{1}+12J_{2}\right)S^{2}$  & 271\\
PFB2a  &  $E_{2,a}=$ & $\left(-2J_{1}-4J_{2}\right)S^{2}$ & -209 \\
PFB2fe  & $E_{2,fe}=$ & $\left(-6J_{1}+12J_{2}\right)S^{2}$  & -406\\
PFB3fm   &$E_{3,fm}=$ & $\left(6J_{1}+8J_{2}+8J_3\right)S^{2}$  & 674\\
PFB3a   &$E_{3,a}=$ & $\left(-6J_{1}+8J_{2}+8J_3\right)S^{2}$  & -35\\
PFB4, FM & $E_{4,fm}=$ & $\left(4J_{1}+12J_{2}+8J_{3}\right)S^{2}$  & 611\\
PFB4a & $E_{4,a}=$ & $\left(-4J_{1}-4J_{2}+8J_{3}\right)S^{2}$  & 101\\
PFB4b & $E_{4,b}=$ & $\left(-4J_{2}-8J_{3}\right)S^{2}$  & 346\\
PFB4c & $E_{4,c}=$ & $\left(-4J_{2}+8J_{3}\right)S^{2}$  & 346\\
PFB5, FM  & $E_{5,fm}=$ & $8\left(J_{1}+J_{2}\right)S^{2}$  & 386\\
PFB5a  & $E_{5,a}=$ & $8\left(-J_{1}+J_{2}\right)S^{2}$  & -530
\end{tabular}\end{ruledtabular}
\end{table}

Using the formulas from the Table~\ref{tab:I}, we find the 
expressions for the
magnetic interactions in the PFB4 chemical configuration 
\begin{eqnarray}
J_3 &=& \left(E_{4,c}-E_{4,b}\right) /16S^2, \label{J3}\\
J_1 &=& \left(E_{4,b}-E_{4,a}\right) /4S^2+4J_3, \label{J1}\\
J_2 &=& \left(E_{4,fm}-E_{4,a}\right) /16S^2-J_1/2. \label{J2}
\end{eqnarray}
Substituting the values of energy differences from the 
Tables~\ref{tab:PFB4}, \ref{tab:I} into these equations, we 
obtain the values of the interactions given in the Table~\ref{tab:II}.
The last row of the table shows the results for fully relaxed lattice.
\cite{Raevski12,rtab}
\begin{table}
\caption{\label{tab:II}
Values of exchange parameters in the PFB4 chemical configuration}
\begin{ruledtabular}    \begin{tabular}{llllll}
$U$, eV & $a$, \AA & $J_1/k_B$, K & $J_2/k_B$, K & $J_3/k_B$, K & $J_2/J_1$ \\
\hline
4 & 4.01 & 92  & 4.3  & $< 0.1$ & 0.046 \\
6 & 4.01 & 75  & 2.0  &  -0.1   & 0.026 \\
4 & 3.95 & 98  & 5.0 &   -0.3   & 0.051 \\
4 & $\sim 3.95$\footnotemark\footnotetext[1]{
Fully relaxed lattice from the calculations in the 
Ref.~\onlinecite{Raevski12}}
 & 113 & 2.4 & $< 0.1$   & 0.021
\end{tabular}
\end{ruledtabular}
\end{table}
Our calculations of the total energies confirm the results of Ref.
\onlinecite{Raevski12}.\cite{rtab} But we find that
the lowest energy for 2nd configuration corresponds to 
{\em ferrimagnetic}
type of ordering PFB2fe.\cite{Kuzian13arX}


%
\begin{table}
\caption{\label{tab:III}
Check of the mapping . DFT energy differences ($U=4$~eV)
are compared with the results for the model,  
Eq.~(\ref{eq:Heff}), which assumes $J_i$ to be {\em independent} on 
the chemical configuration.
}
\begin{ruledtabular}  \begin{tabular}{crcr}
$\Delta E/S^{2}$, meV & DFT & Model & Value\\
\hline
$(E_{5,fm}-E_{5,a})/S^{2}$ & 146.6 & $16J_1$ & 156.5 \\
$(E_{3,fm}-E_{3,a})/S^{2}$ & 113.4 & $12J_1$ & 117.3 \\
$(E_{2,fm}-E_{2,fe})/S^{2}$ & 108.3 & $12J_1$ & 117.3 \\
$(E_{2,fm}-E_{2,a})/S^{2}$ & 76.7 & $8J_1+16J_2$ & 81.6 \\
$(E_{2,a}-E_{2,fe})/S^{2}$ & 31.6 & $4J_1-16J_2$ & 35.8 \\
$(E_{1,fm}-E_{1,a})/S^{2}$ & 5.3 & $16J_2+32J_3$ & 3.3 \\
$(E_{1,fm}-E_{1,b})/S^{2}$ & 79.9 & $8J_1+16J_2$ & 81.6 \\
$(E_{0,fm}-E_{0,a})/S^{2}$ & 11.1 & $32J_2$ & 6.7 
\end{tabular}\end{ruledtabular}
  \end{table}

The Table\ \ref{tab:III} shows the results of the check
of the quality of our mapping of LSDA+$U$ on the
Heisenberg model (\ref{eq:Heff}).

\section{Discussion}

\subsection{Superexchange interaction}

Our calculations strongly suggest that the magnetism of 
AFe$_{1/2}$M$_{1/2}$O$_{3}$
systems may be described by the Heisenberg model on the lattice which
is obtained from the simple cubic lattice by removing half of its
sites, the nearest-neighbor interaction $J_1$ being dominant. 
\begin{figure}[hbt]
\includegraphics[width=0.33\columnwidth]{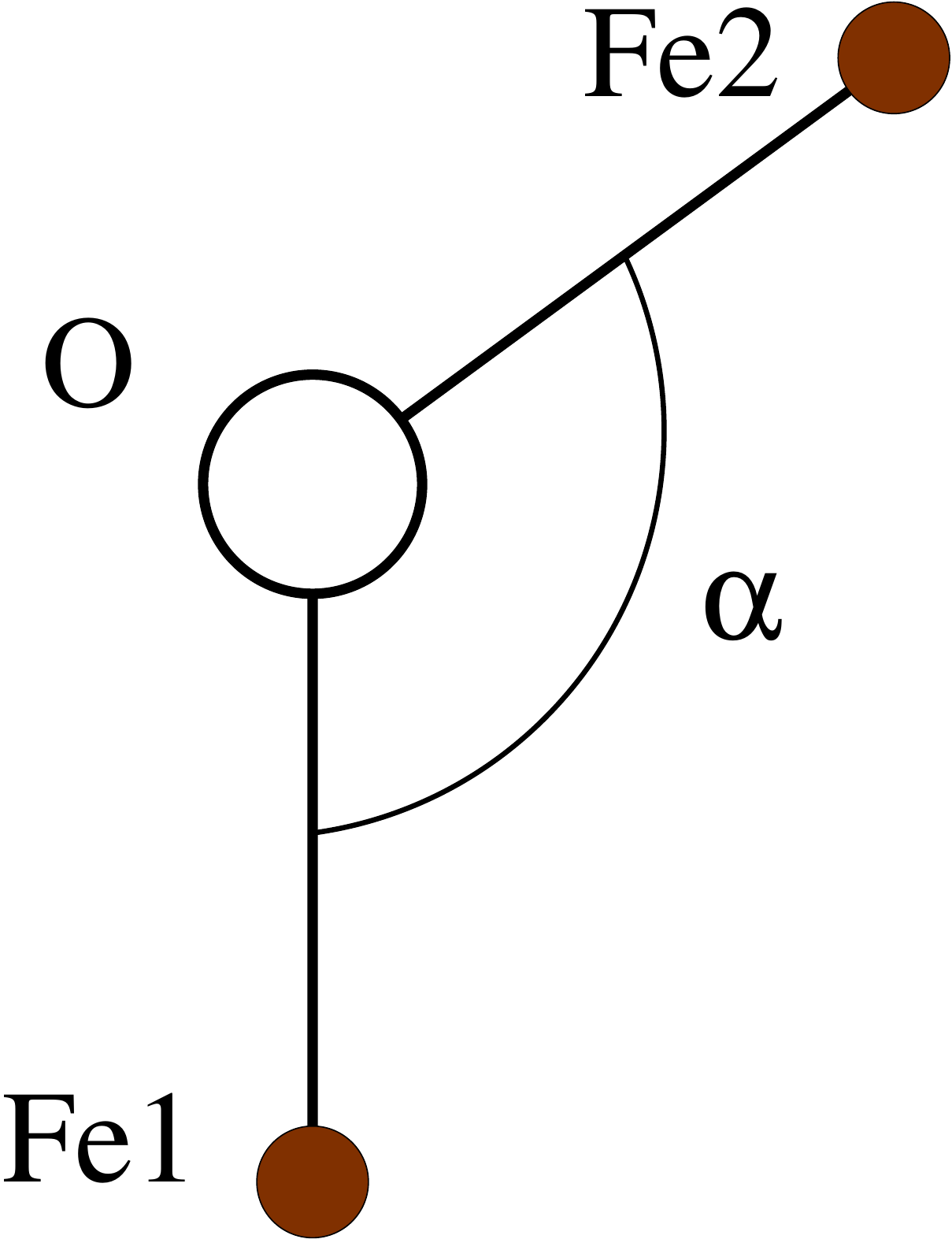} 
\caption{(Color online)
The geometry of Fe$_1$-O-Fe$_2$ superexchange path.
}
\label{Fe2O}
\end{figure}

The dominance of $J_1$ is an expected results. The magnetic interactions 
between Fe$^{3+}$ ions are due to the superexchange 
mechanism\cite{Anderson63}, which has a local 
nature for $3d$-metal compounds\cite{Larson88,Larson89}.The ion Fe$^{3+}$
has $d^5$ electronic configuration. For this configuration, 
the fourth-order many-body perturbation theory expression for the 
superexchange  via a single intervening oxygen ion (Fig.~\ref{Fe2O})
may be written\cite{Kuzian10,Kuzian11} in a simple form 
(see Appendix~\ref{AppB} for the derivation)
\begin{eqnarray}
J_{\alpha} &\approx & KV_{pd\sigma,1}^{2}V_{pd\sigma,2}^{2}
 \left(0.475+0.617\cos^{2}\alpha\right), \label{JFeFe} \\
 &=& J_{180}\left(0.475+0.617\cos^{2}\alpha\right)/1.092 \label{Jpi}\\
 &=& J_{180}\cos^{2}\alpha + J_{90}\sin^{2}\alpha \label{JSaw}
\end{eqnarray}
where $\alpha$ is the Fe-O-Fe bond angle; 
$K$ is given by Eq.(\ref{Kden}), it does not depend on the bond geometry, 
$V_{pd\sigma,i}$ are the Slater-Koster\cite{SK} parameters
for the electron hopping integrals between Fe and O ions, which
depend only on the Fe-O bond-lengths. 

The dependence of the Fe-O-Fe
superexchange on the  square of the bond angle cosine $\cos^{2}\alpha$
was established for the orthoferrites RFeO$_{3}$ in the 
Ref.~\onlinecite{Boekema72}
in the form given by Eq.(\ref{JSaw}). For the RFeO$_{3}$ family, the 
bond angle varies between 157$^\circ$ in LaFe0$_3$ to 
142$^\circ$ in LuFe0$_3$, the Fe-O bond-length being approximately
constant $d \approx 2.01$~\AA. Substituting the LuFe0$_3$ parameters
$\cos^{2}\alpha \approx 0.618$ and $J/k_B \approx 48.4\pm 2$~K
into Eq.(\ref{Jpi}) we find for $J_{180}/k_B \approx 62$~K, which is 
comparable with our $J_1$ value calculated for $U=6$. 
The assumption (ii) from the Introduction is thus confirmed.

Our formula (\ref{JFeFe}) shows also that the Fe-O-Fe
superexchange depends on the Fe-O bond lengths $R_i$. The hopping
integrals $V_{pd\sigma,i}(R_i)$ decrease with the increase of 
the bond length.\cite{Harrison} 
This means that the superexchange should decrease with the increase of
lattice parameter if the bond angle remains constant. 
The results shown in the Table~\ref{tab:II} follows this tendency.

We may compare our results also with the Ref.~\onlinecite{Lampis04}, 
where 
the values for $J_1$, $J_2$, and fourth-neighbor $J_4$ exchanges
were found for PFB0[=$\frac{1}{2}(111)$] and PFB5[=$\frac{1}{2}(100)$] 
configurations (Fig.~\ref{mstr}). If we express the results
from the Table III of Ref.~\onlinecite{Lampis04} in our notations, we
obtain $-2J_1^s/k_B=J_2/k_B \approx 0.9$~K, and 
$-2J_2^s/k_B=J_4/k_B \approx 2.8$~K for PFB0, 
and  $-2J_1^s/k_B=J_1/k_B \approx 42$~K, 
$-2J_d/k_B=J_2/k_B \approx 0.5$~K, 
$-2J_2^s/k_B=J_4/k_B \approx 2.8$~K for PFB5. The results of
Ref.~\onlinecite{Lampis04} confirm the dominance of $J_1$
nearest-neighbor Fe-O-Fe interaction. The absolute value of the 
interaction is smaller, but we should take into account that the authors
of the Ref.~\onlinecite{Lampis04} have used $U=9$~eV value in those 
calculations. Note that they obtained 
$J_1/k_B \approx 50$~K for LaFeO$_3$, which is slightly
smaller than experimental value\cite{Eibschuetz67} 59~K derived from
$T_N=740$~K using high-temperature expansion.

\subsection{Collective magnetic properties}

The way how the half of sites of simple cubic lattice are 
occupied by the interactiong Fe spins determine the 
magnetic
properties of the system. In this work, we model the disorderd system
by 2$\times$2$\times$2 supercell periodic lattice.
If we take into account only nearest-neighbor interaction $J_{1}$,
then magnetic ions form three dimensional lattice only in PFB2 and
PFB3 configurations (Fig.~\ref{mstr}). Thus, only these configurations 
may possess a
magnetic long-range order at non-sero temperature. Ohter configurations
have lower dimensionalities and thus have no ordering at finite
temperatures. Actually, small next-nearest neighbor interactions 
(like $J_2$, $J_3$) will ensure the ordering, but the temperature 
will be substantially lower (see below the consideration of PFB5
structure in the subsection~\ref{mag5}).

The simplest molecular field approach gives for the ferrimagnetic
ordering temperature (see  Appendix~\ref{AppC} for the details) of PFB2fe
configuration
\begin{equation}
T_{2fe}=J_{1}2\sqrt{3}\frac{S(S+1)}{3k_{B}}
\approx 10.1J_1,
\label{eq:T2fe}
\end{equation}
 and for the antiferromagnetic ordering temperature of PFB3a configuration
\begin{equation}
T_{3a}=J_{1}2\frac{S(S+1)}{3k_{B}}\sqrt{\frac{3+\sqrt{5}}{2}} 
\approx 9.44J_1.
\label{eq:T3a}\end{equation}
 Substitution of the calculated $J_{1}$ value gives 
 $T_{2fe}\approx 933$(758)~K,
$T_{3a}\approx 872$(708)~K for U=4(6)~eV. We should have in mind that the 
molecular field theory overestimates
the transition temperature by the factor $\sim 1.5$ for cubic lattices
and this factor may increase for the structures with the number of
neighbors less than 6. 
Indeed, a more accurate estimate may be derived 
from the high-temperature 
expansion of the magnetic susceptibility $\chi$.
In the Ref.~\onlinecite{Kuzian13arX},
we have applied the method and the program package for the eighth-order 
high-temperature expansion
for a general Heisenberg model with up to four 
different exchange parameters $J_1, J_2, J_3, J_4$ presented
recently in the Ref.\ \onlinecite{Schmidt11,hte}. 
The temperature for the transition into  ferrimagnetically ordered phase 
$T_{fe,HTE}$ is defined as the point where $\chi ^{-1}(T_{fe})=0$. 
We have obtained $T_{fe,HTE}\approx 5.6J_1 \approx 517(420)$ for 
 U=4(6)~eV (see the details in the Ref.~\onlinecite{Kuzian13arX}).

The ferrimagnetism in PFB2 chemical order has rather unusual nature. 
In many cases, the ferrimagnetism is due to different spin values of
ions occupying different antiferromagnetically coupled 
magnetic sublattices. Another 
possibility is realized e.g. in the yttrium iron garnet Y$_3$Fe$_5$O$_{12}$
and related compounds. There, all Fe ions have equal spins $S=5/2$,
but the lattice has two kinds of Fe positions, and the number of Fe
sites in antiparallel sublattices is 
different\cite{Geller57,Cherepanov93}. So, the ferrimagnetism may have
a purely geometrical origin\cite{Lieb62}. This is the case for the 
magnetic ground state of the PFB2 chemicaly ordered 
lattice.\cite{Kuzian13arX}

We understand that 2$\times$2$\times$2 supercell periodic lattice
is a rather poor approximation to the disorded system. Nevertheless,
it is instructive to estimate the probabilities to find different
chemical configurations PFBn (see Fig.~\ref{mstr} and Fig.~3 of the
Ref.~\onlinecite{Raevski12}). If the system is totally disordered, i.e. 
if Fe and  M ions randomly occupy B sites
of the perovskite lattice (assumption (i) of the Introduction),
we have $C_{8}^{4}=70$ ways to ditribute
Fe ions over 8 vertex of the cube, every configuration being equivalent
to one of that depicted on the Fig.~\ref{mstr} We will meet 2 times
the configuration PFB0, 6 times PFB1 and PFB5 configurations, 8 times
PFB2, and 24 times PFB3 and PFB4 configurations 
$2+2\times6+8+2\times24=70$.
So, in the case of random distribution, the probability to meet PFB2
configuration is $P_{2}=8/70\approx 0.11$, and to meet PFB3 configuration
is $P_{3}=24/70\approx 0.34$. Thus the magnetic properties of an
AFe$_{1/2}$M$_{1/2}$O$_{3}$
compound will be dominated by PFB3 configuration. So, within our simple 
model
of disoder the transition temperature would be several hundreds K. As we 
have
mentioned in the introduction, more sophisticated treatment of the disorder
results in $T_N\sim 300$~K.\cite{Stinchcombe79,Kumar81}
Evidently, the assumption (i) is in contradiction with the observed 
values of transition and Curi-Weiss temperatures.

The distribution of Fe and M ions over B sites depends on the
ratio of ionic radii of Fe and M metal ions, the growth condition
of the sample etc.  When the radius of M$^{5+}$ ion is larger than that 
of Fe$^{3+}$,
the ordered PFB0 configuration becomes most probable\cite{Raevski12}. 
This is often the case
for M=Sb.\cite{Battle95,Raevskii02} 
For such an 1:1 ordered systems, magnetic Fe$^{3+}$
ions form regular face centered cubic sublattice with interaction $J_2$
between nearest spins in the sublattice. The Curie-Weiss temperature
is $\Theta _{CW,0}=  4S(S+1)J_2/k_B$. The magnetic ground state of such
Heisenberg lattice is so called I-type order, which is denoted as PFB0a
on the Fig.\ \ref{mstr}. The transition temperature was studied in
Ref.\ \onlinecite{Pirnie66} using high temperature series expansion. It
occurs to be spin independent and equals 
$T_I \approx  -\Theta _{CW,0}/5.76$.
\begin{table}
\caption{\label{tab:4} The Curie-Weiss  $\Theta _{CW,0}$ and calculated
transition $T_{I}=-\Theta _{CW,0}/5.76$ temperatures
for 1:1 ordered systems. DFT calculations and experimental results.
The temperature of observed susceptibility anomaly $T_{max}$ is shown for 
two 
compounds.}
\begin{ruledtabular} \begin{tabular}{cccc}
$U$, eV  & $\Theta _{CW,0}$, K & $T_{I}$, K & $T_{max}$, K\\
\hline
4 &  -151 & 26 & \\
6 &  -70 &  12 & \\
Sr(Fe$_{1/2}$Sb$_{1/2}$)O$_{3}$\footnotemark\footnotetext[1]{
Reference \onlinecite{Battle95a}}
& -221 & 38 & 36\\
Ca(Fe$_{1/2}$Sb$_{1/2}$)O$_{3}$\footnotemark\footnotetext[2]{
Reference \onlinecite{Battle95}}
 & -89 & 15 & 17
   \end{tabular}
%
\end{ruledtabular}
\end{table}
The Table~\ref{tab:4} compares the calculated values of $T_{I}$ with 
the temperature $T_{max}$ of the magnetic susceptibility anomaly  
observed in 1:1 ordered AFe$_{1/2}$M$_{1/2}$O$_{3}$ compounds.

\subsection{
Magnetism of the PbFe$_{1/2}$M$_{1/2}$O$_{3}$ compounds}\label{mag5}

The total energies of various chemical configurations 
(see Fig.~\ref{mstr}) of Fe in
a 2$\times$2$\times$2 supercell of PbFe$_{1/2}$M$_{1/2}$O$_{3}$ 
(M=Nb, Ta, Sb) were
calculated in Ref.~\onlinecite{Raevski12} using the LSDA+$U$ functional.
For the PFN and PbFe$_{1/2}$Ta$_{1/2}$O$_{3}$ (M=Nb, Ta) 
compounds, the layered PFB5a 
configuration has the lowest energy, in contrast to 
PbFe$_{1/2}$Sb$_{1/2}$O$_{3}$,
where the PFB0 1:1 chemically ordered configuration is most favorable
\cite{rtab} (see also Table~\ref{tab:I}).  
The PFN and PbFe$_{1/2}$Ta$_{1/2}$O$_{3}$ 
compounds are especially interesting
because they are multiferroics and exhibit ferroelectric transition 
($T_C \approx $380,
270~K for M=Nb, Ta) in addition to antiferromagnetic transition. 
As we have mentioned in the 
Introduction, the peculiarity of magnetic properties of these two 
compounds is that
those N\'{e}el temperature $T_N \sim 150$~K is much higher than the 
transition 
temperature for other double perovskites. A layered Heisenberg model 
with the 
nearest neighbor interaction $J_1$ within the layer and an interlayer 
interaction 
$J_{\perp}$ was  thoroughly studied in the past 
(see Ref.\ \onlinecite{Junger09} 
and references therein). It was established that the transition 
temperature has 
logarithmic dependence on $J_{\perp}/J_1$ ratio
\begin{equation}
\frac{T_N}{T_{N,sc}} \approx \frac{1}{1-k\ln (J_{\perp}/J_1)} ,
\label{TNlayer}\end{equation}
where $T_{N,sc}$ is the transition temperature for the G-type 
antiferromagnetic ordering of 
simple cubic  lattice ($J_{\perp} = J_1$), and $k\approx 1/3$. The 
equation 
(\ref{TNlayer}) was found to work in the wide range of values 
$0.001\leq J_{\perp}/J_1 \leq 1$ \cite{Yasuda05,Junger09}, it gives
$T_N/T_{N,sc} \approx $  0.30, 0.39, 0.57 for $J_{\perp}/J_1 =$ 0.001, 
0.01, 0.1 
respectively. Taking $T_{N,sc} \sim 600$~K, we obtain reasonable values 
for $T_N \approx $ 
180, 234, 342, respectively,
if we assume that PFN and PbFe$_{1/2}$Ta$_{1/2}$O$_{3}$ 
have the totally ordered layered 
structure.  

In reality, both compounds are disordered and the results of the 
supercell calculations
indicate only what kind of {\em short-range} chemical order is 
more favorable. 
Below $T_N$, the neutron diffraction 
studies\cite{Pietrzak81,Ivanov00,Rotaru09} reveal a G-type 
antiferromagnetic
order  with magnetic moments $\mu \approx 2.8\mu _B$ sitting at 
{\em every} site of the simple cubic lattice. It is clear that this is
an averaged picture with a "half of Fe$^{3+}$ ion" in every site of 
B-sublattice of the structure. The value of $\mu$ is about 
half of the value expected for Fe$^{3+}$ ion $\mu _{Fe}=5\mu _B$.

In contrast to neutron diffraction, local probe methods such as Nuclear 
Magnetic Resonance (NMR) and Mossbauer spectroscopy provide local 
structure information. In this respect, we can mention $^{17}$O NMR 
data\cite{Blinc08} which may confirm our theoretical prediction that 
PFB5 configuration gives major contribution to the 
antiferromagnetic ground state of 
PFN or PbFe$_{1/2}$Ta$_{1/2}$O$_{3}$. 
Indeed, $^{17}$O NMR spectrum consists of two distinct 
components: narrow and very broad lines. One can see from 
Fig.~\ref{scell} that 
each O ion connects only two cations forming three different pathways 
along $\langle 100 \rangle$ cubic directions: Fe-O-Fe, Fe-O-Nb, and Nb-O-Nb. The first 
two configurations are responsible for the broad component in $^{17}$O 
NMR 
spectrum as O nucleus is closely adjacent to the magnetic Fe$^{3+}$ ion. 
The last configuration does not contain magnetic ions therefore is 
responsible for the narrow component in NMR spectrum. Among all chemical 
configurations shown in Fig.~\ref{mstr}, the PFB5 configuration has a 
largest number of nonmagnetic Nb-O-Nb chains. Assuming, for example, 
random distribution of Fe and Nb ions we have relative weight of the 
Nb-O-Nb pathways only 0.19, while the NMR data predicts 2-2.5 times 
larger value. This suggests that the layered PFB5 chemical configuration 
can dominate among other chemical ordering. The non random distribution 
of magnetic and nonmagnetic cations in PFN is also supported by $^{93}$Nb 
NMR measurements.\cite{Laguta10} The NMR data have been interpreted in 
a model which assumes existence of Fe rich, Nb poor and Fe poor, 
Nb rich regions in PFN.

The Table~\ref{tab:I} shows that the PFB2fe configuration has the
total energy,
which is close to the lowest PFB5 configuration. 
This is the case also for PbFe$_{1/2}$Ta$_{1/2}$O$_{3}$ and 
PbFe$_{1/2}$Sb$_{1/2}$O$_{3}$.\cite{rtab}
A sample of a disordered double perovskite compound may contain some 
regions with PFB2 chemical order.
In the ground state, such a region possess the moment
$\mu _g = N_c\mu _c$, where $N_c$ is the number of supercells in the 
region, $\mu _c=10\mu _B$ is the moment of the supercell. Large moment of the 
region will persist for $T<T_{2fe}$. 
Therefore, it can not be excluded that such regions exist in the
systems 
PbFe$_{1/2}$Ta$_{1/2}$O$_{3}$/PbZr$_x$Ti$_{1-x}$O$_{3}$\cite{Sanchez11,Evans13}, 
PFN/PbZr$_x$Ti$_{1-x}$O$_{3}$\cite{Sanchez13}, 
and PbFe$_{2/3}$W$_{1/3}$O$_3$/PbZr$_x$Ti$_{1-x}$O$_{3}$\cite{Kumar09}
and are responsible for for large room-temperature magnetic response
and magnetoelectric coupling in spite that 
the long-range magnetic order establishes 
far below the room temperatures.

\section{Conclusion}

Based on LSDA+$U$ calculations, 
we have found that AFe$_{1/2}$M$_{1/2}$O$_{3}$ 
double perovskite compounds may be described by antiferromagnetic 
$J_1-J_2$
Heisenberg model on the lattice, which
is obtained from the simple cubic lattice by removing half of its sites. 
The dominant magnetic interaction is Fe-O-Fe superexchange $J_1$ between
Fe$^{3+}$ (S=5/2) ions occupying nearest-neighbor positions
within the B-sublattice of ABO$_3$ perovskite structure.
The next-nearest-neighbor interaction $J_2$ which
corresponds to sites separated by the face diagonal 
of perovskite unit cell is much smaller.
The estimated 
values of the exchange paramters are close to the values reported for 
Fe-based perovskites RFeO$_3$, where all octahedral sites are 
occupied by Fe ions. The distribution of Fe$^{3+}$ 
and M$^{5+}$ ions over B sites of perovskite lattice determine the magnetic 
properties of the double perovskites. Our results suggest that the 
distribution is 
{\em not} random. The typical value of the magnetic 
transition temperature  $T_N \sim 25$~K 
in most of the paraelectric double perovskite compounds
allows to conclude that
the probability to find there a nearest-neighbor pair of Fe 
(interacting with $J_1$ exchange value) is suppressed 
compared the probability to find the next-nearest pair,
and the magnetic energy scale is determined by $J_2$.
In accord with Ref.\ \onlinecite{Raevski12}, we argue  that two 
multiferroic compounds PbFe$_{1/2}$Nb$_{1/2}$O$_{3}$
and PbFe$_{1/2}$Ta$_{1/2}$O$_{3}$ 
($T_N \sim 150$~K) have predominantly layered PFB5 (see Fig.~\ref{mstr})
short-range ordering where the B-sublattice is formed by alternating
Fe and M(=Nb or Ta) planes.

We have also found that Fe ion in double perovskites may form a
sub-nano-sized superstructure (PFB2 chemical order in the Fig.~\ref{mstr})
having  the room temperature {\em ferrimagnetic} order.
Such ferrimagnetism of geometrical origin\cite{Lieb62} may represent an 
interesting alternative to the room-temperature ferromagnetism
in wide-gap semiconductors, which is in the focus of recent studies.
Formation of the PFB2 superstructure in ferroelectric double perovskites
will lead to the room-temperature multiferroism where ferroelectric and 
ferrimagnetic type order can coexist, at least at a nanoscale level. 
Recent observations of the room-temperature 
multiferroism in complex systems
on the base of the double 
perovskites\cite{Sanchez11,Evans13,Sanchez13,Kumar09}
are possibly provided by nano-regions of the ferrimagnetic 
PFB2 superstructure
rather than by simple local clustering of Fe ions as this will lead only 
to increase of Neel  temperature. .

\begin{acknowledgments}
The authors thank M.\ D.\ Kuz'min
 for very useful discussions, 
S.\ A.\ Prosandeev for providing the details on numerical calculations
in Ref.\onlinecite{Raevski12}, 
U.~Nitzsche and K.~Koepernik for technical assistance,
and the IFW Dresden (Germany) that we could use their computer facilities.
The project GACR 13-11473S is acknowledged.
\end{acknowledgments}

\appendix

\section{Detail of the LSDA+$U$ calculations}\label{AppA}

\begin{figure}
\includegraphics[width=\columnwidth]{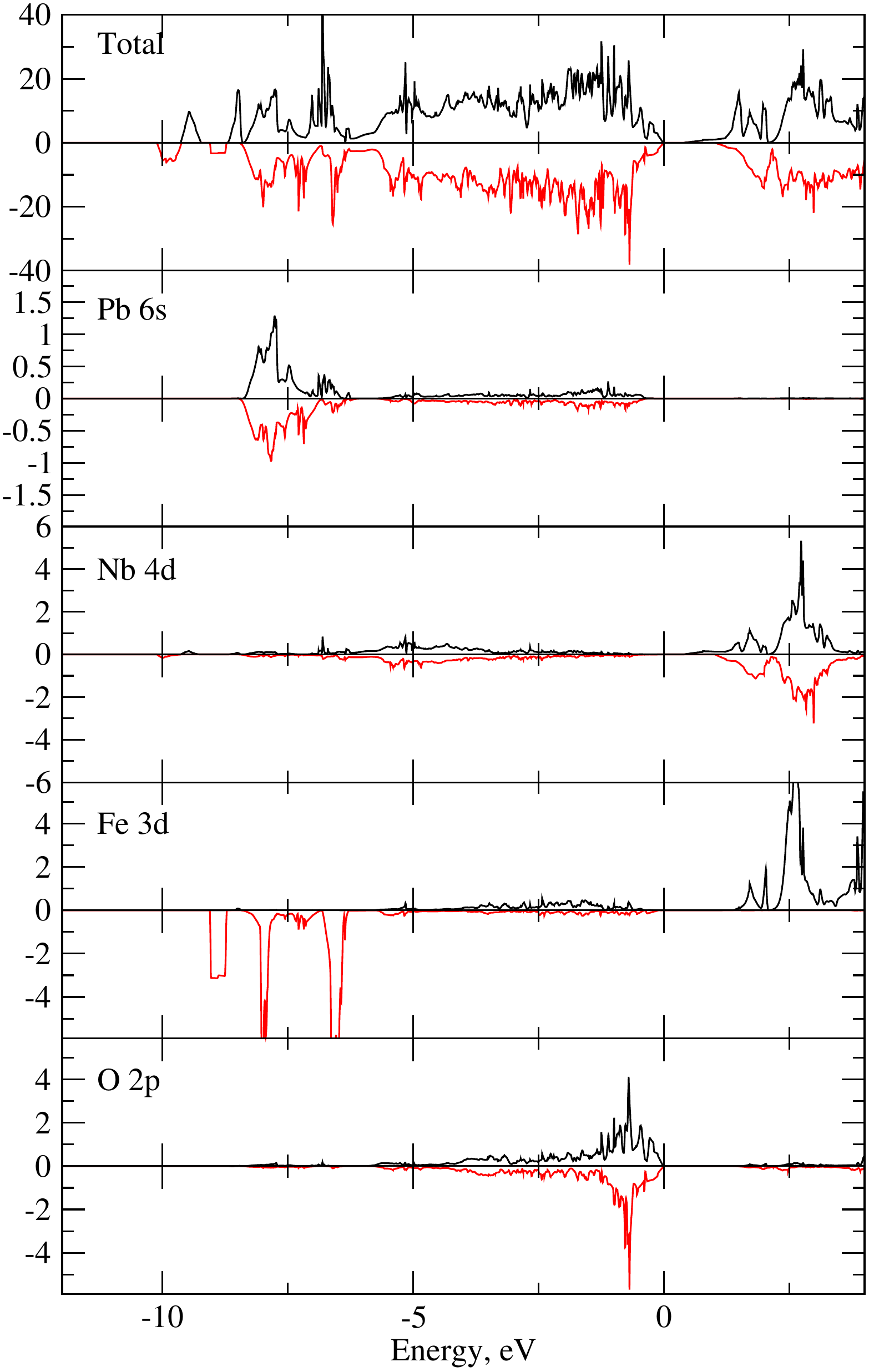} 
\caption{(Color online)
Spin resolved total density of states (upper panel)
for PFB4a structure, $U=6$~eV. Other panels show
representative densities of states 
projected onto the basis functions (one for every ion sort), which 
maximally contribute to the total density of state near the Fermi level.
}
\label{dos}
\end{figure}

Total and projected densities of states are shown in the Fig.~\ref{dos}.
As expected, the largest spin splitting occurs for Fe $3d$-states.
The Table~\ref{tab:A1} shows the values of magnetic moments localized 
on the Fe ions. They are close to the isolated Fe$^{3+}$ ion value 
5$\mu _B$. In the groundstate PFB4a configuration, the polarization 
do not exceed 0.03$\mu _B$ for oxygen ions,
and 0.04$\mu _B$ for Nb ions. In the ferromagnetic state, the 
polarization of some oxygens and Nb ions reaches 0.14$\mu _B$ and 
0.12$\mu _B$ respectively.
\begin{table}
\caption{\label{tab:A1} The magnetic moments 
(in the units of Bohr magneton $\mu _B$)
localized on different Fe ions of PFB4 chemical 
configuration in various magneitc states
depicted in Figs.~\ref{mstr}, \ref{mstr2}.
}
\begin{ruledtabular} \begin{tabular}{cccrrrr}
 &$U$, eV & $a$, \AA & Fe1 & Fe2 & Fe3 & Fe4\\
\hline
a & 4 & 4.01 &4.45 & 4.45 &-4.39 &-4.52 \\
      & 6 & 4.01 &4.77 & 4.77 &-4.71 &-4.85 \\
      & 4 & 3.95 &4.49 & 4.49 &-4.42 &-4.56 \\
  & 4 &$\sim 3.95$\footnotemark\footnotetext[1]{
Fully relaxed lattice from the calculations in the 
Ref.~\onlinecite{Raevski12}}
  &4.43 &4.43 &-4.35&-4.45 \\
\hline
b & 4  &  4.01    &4.49 &-4.45 & 4.43 &-4.51 \\
      & 6 &   4.01    &4.79 &-4.77 & 4.73 &-4.85 \\
   & 4 & 3.95 & 4.52 & -4.49 & 4.45 &-4.55 \\
  & 4 &$\sim 3.95$\footnotemark[1]&4.49 &-4.43 &4.41 &-4.45 \\
\hline
FM & 4 & 4.01 & 4.50 & 4.50 & 4.48 & 4.50 \\
   & 6 & 4.01 & 4.80 & 4.80 & 4.75 & 4.85 \\
   & 4 & 3.95 & 4.53 & 4.53 & 4.49 & 4.54 \\
 & 4 & $\sim 3.95$\footnotemark[1] & 4.50& 4.49 &4.47 &4.45
\end{tabular}
\end{ruledtabular}
\end{table}
We see that the localized moment description of the magnetism in PFN 
by the model Hamiltonian (\ref{eq:Heff}) is adequate.

\section{Three center cation-anion-cation model}\label{AppB}

Here we will calculate the superexchange for the case when it is mediated
by one anion where the CF splitting will be neglected. 
Then we choose
the coordinate system having the anion in the origin, and the vector
radii of the cations are 
\begin{eqnarray*}
\mathbf{R}_{1} & = & \left(0,0,-1\right)R_{1}\ ,\\
\mathbf{R}_{2} & = & \left(\sin\alpha,0,-\cos\alpha\right)R_{2}\ ,
\end{eqnarray*}
 where $\alpha$ is the angle between bonds (Fig.~\ref{Fe2O}). 
A general fourth-order
many-body perturbation theory expression for the superexchange between
two ions in $d^{5}$ configuration reads (cf. Eqs. (9),(10) of the
Ref.~\onlinecite{Kuzian11})
\begin{equation}
J=-\frac{1}{2S^{2}\Delta_{eff}^{2}}
\left(\frac{r^{2}}{U_{eff}}+\frac{2}{2\Delta_{eff}+
U_{p}}\right)E_{\beta \beta} 
\label{eq:Jdd}
\end{equation}
 where 
\begin{table*}[htb]
\caption{Hoppings $t_{2m^{\prime}n^{\prime}}$ between $t_{2g}$
orbitals and ligand $p$-functions}\label{Flo:tt2g}
\begin{ruledtabular}\begin{tabular}{c|ccc}
n\textbackslash{}m  & $xy$  & $yz$  & $zx$\tabularnewline
\hline 
$x$  & 0  & 0  & $-\cos\alpha\left[\sqrt{3}\sin^{2}\alpha V_{pd\sigma}+
\left(1-2\sin^{2}\alpha\right)V_{pd\pi}\right]$\tabularnewline
$y$  & $\sin\alpha V_{pd\pi}$  & -$\cos\alpha V_{pd\pi}$  & 0\tabularnewline
$z$  & 0  & 0  & -$\sin\alpha\left[\sqrt{3}\cos^{2}\alpha V_{pd\sigma}+
\left(1-2\cos^{2}\alpha\right)V_{pd\pi}\right]$\tabularnewline
\end{tabular}\end{ruledtabular}
\end{table*}
\begin{table*}[htb]
\caption{Hoppings $t_{2m^{\prime}n^{\prime}}$ between $e_{g}$ 
orbitals and ligand $p$-functions}
\label{Flo:teg}
\begin{ruledtabular}\begin{tabular}{c|cc}
n\textbackslash{}m  & $x^{2}-y^{2}$  & $z^{2}$\tabularnewline
\hline 
$x$  & $\frac{\sqrt{3}}{2}\sin^{3}\alpha V_{pd\sigma}+
\sin\alpha\left(1-\sin^{2}\alpha\right)V_{pd\pi}$  
& $\sin\alpha\left[\left(\cos^{2}\alpha-\frac{\sin^{2}\alpha}{2}\right)
V_{pd\sigma}-\sqrt{3}\cos^{2}\alpha V_{pd\pi}\right]$\tabularnewline
$y$  & 0  & 0\tabularnewline
$z$  & $-\cos\alpha\left[\frac{\sqrt{3}}{2}\sin^{2}\alpha V_{pd\sigma}-
\sin^{2}\alpha V_{pd\pi}\right]$  & $-\cos\alpha
\left[\left(\cos^{2}\alpha-\frac{\sin^{2}\alpha}{2}\right)
V_{pd\sigma}+\sqrt{3}\sin^{2}\alpha V_{pd\pi}\right]$\tabularnewline
\end{tabular}\end{ruledtabular}
\end{table*}
\begin{eqnarray}
E_{\beta\beta} & = & \sum_{m,m^{\prime},n,n^{\prime}}
t_{1,m,\beta ,n}t_{2,m^{\prime},\beta ,n}t_{1,m,\beta,n^{\prime}}
t_{2,m^{\prime},\beta ,n^{\prime}}, \label{eq:Ebb}\\
U_{eff} & = & U_{d}+4J_{H},\label{eq:Ueff}\\
\Delta_{eff} & = & \Delta+28J_{H}/9. \label{eq:Deff}
\end{eqnarray}
The $d$-ions are assumed to be in the high-spin state ($S=5/2$),  
$U_{d}$ ($U_p$) is the the Coulomb repulsion between two fermions on the
same $d$-($p$-)orbital, $J_{H}\equiv\frac{5}{2}B+C$ is the Hund exchange
in the $d$-shell expressed in terms of Racah parameters, 
and $\Delta$ is the charge transfer energy
(see Ref.~\onlinecite{Kuzian11} for the discussion of the approximations
behind the Eq.(\ref{eq:Jdd}), and the description of the 
realistic many-body $p-d$ Hamiltonian). 
According to Harrison model\cite{Harrison}, the
hopping $t_{\alpha,m,\beta,n}$ between $m$-th $d$-function of metal
ion $\alpha=1,2$ and $n$-th $p$-function of ligand $\beta$ 
is expressed via direction cosines $l,m,n$ of the
vector $\mathbf{R}_{\beta}-\mathbf{R}_{\alpha}$, and two 
Slater-Koster\cite{SK}
parameters $V_{pd\sigma}(R),V_{pd\pi}(R)$ , which depend on sorts
of metal ion and on the distance 
$R=\left|\mathbf{R}_{\beta}-\mathbf{R}_{\alpha}\right|$;
$r\approx0.8$ is a reduction factor that is caused by dependence
of the hoppings on the number of $3d$ electrons. 

In the case of single ligand, the index $\beta$ may be droped, and
it is convenient to write 
\begin{eqnarray*}
E & \equiv & \sum_{n,n^{\prime}}\sum_{m,m^{\prime}}t_{1,m,n}
t_{2,m^{\prime},n}t_{1,m,n^{\prime}}t_{2,m^{\prime},n^{\prime}}\ ,\\
 & = & \sum_{n,n^{\prime}}\sum_{m}t_{1,m,n}t_{1,m,n^{\prime}}
 \sum_{m^{\prime}}t_{2,m^{\prime},n}t_{2,m^{\prime},n^{\prime}}\ .
\end{eqnarray*}

The Slater-Koster table\cite{SK,Harrison} gives for the first 
transition metal-anion pair
\begin{eqnarray*}
t_{1zx,x} & = & t_{1zy,y}=-V_{pd\pi,1}\ ,\\
t_{1z^{2},z} & = & -V_{pd\sigma,1},
\end{eqnarray*}
 other hoppings are zero. So
\begin{eqnarray*}
\sum_{m}t_{1,m,n}t_{1,m,n^{\prime}} & = & \delta_{nn^{\prime}}
\sum_{m}t_{1,m,n}^{2}\equiv\delta_{nn^{\prime}}T_{1n}\\
T_{1x} & = & T_{1y}=V_{pd\pi,1}^{2}\ ,\ T_{1z}=V_{pd\sigma,1}^{2}\ ,
\end{eqnarray*}
 then
\begin{eqnarray*}
E & = & \sum_{n,n^{\prime}}\delta_{nn^{\prime}}T_{1n}\sum_{m^{\prime}}
t_{2,m^{\prime},n}t_{2,m^{\prime},n^{\prime}}=\sum_{n}T_{1n}T_{2n}\ ,\\
T_{2n} & \equiv & \sum_{m^{\prime}}t_{2,m^{\prime},n}^{2}\ .
\end{eqnarray*}

For the second anion-TMI pair the hoppings are given in tables
\ref{Flo:tt2g},\ref{Flo:teg}. This gives us
\begin{eqnarray*}
T_{2x} & = & \sin^{2}\alpha V_{pd\sigma,2}^{2}+
\cos^{2}\alpha V_{pd\pi,2}^{2}\ ,\\
T_{2y} & = & V_{pd\pi,2}^{2}\ ,\\
T_{2z} & = & \cos^{2}\alpha V_{pd\sigma,2}^{2}+
\sin^{2}\alpha V_{pd\pi,2}^{2}.
\end{eqnarray*}
 And we obtain 
\begin{eqnarray}
E & = & V_{pd\pi,1}^{2}\left(T_{2x}+T_{2y}\right)+V_{pd\sigma,1}^{2}
T_{2z}\nonumber \\
 & = & V_{pd\pi,1}^{2}\left[\sin^{2}\alpha V_{pd\sigma,2}^{2}+
 \left(1+\cos^{2}\alpha\right)V_{pd\pi,2}^{2}\right]\nonumber \\
 & + & V_{pd\sigma,1}^{2}\left(\cos^{2}\alpha V_{pd\sigma,2}^{2}+
 \sin^{2}\alpha V_{pd\pi,2}^{2}\right)\label{eq:E}\\
 & = & V_{pd\sigma,1}^{2}V_{pd\sigma,2}^{2}\frac{1+2\tau^{2}+
 \left(\tau^{2}-1\right)^{2}\cos^{2}\alpha}{\tau^{4}}\nonumber \\
 & \approx & V_{pd\sigma,1}^{2}V_{pd\sigma,2}^{2}
 \left(0.475+0.617\cos^{2}\alpha\right).\label{eq:E12}
\end{eqnarray}
 In the last equality we have introduced the ratio 
 $\tau\equiv V_{pd\sigma}/V_{pd\pi}\approx-2.16$.\cite{Harrison}

Finally, we obtain the Eq.(\ref{JFeFe}) of the main text with
\begin{equation}
K = \frac{r^{2}}{U_{eff}}+\frac{2}{2\Delta_{eff}+
U_{p}} \label{Kden}
\end{equation}

\section{Transition temperature}\label{AppC}

Here we give the derivation of Eqs.(\ref{eq:T2fe}),(\ref{eq:T3a})
for transition temperatures within the molecular field approximation
(see e.g. Ref.\onlinecite{Morrish}).

In the PFB2fe configuration we have two sublattices: Fe1 with spin up
and Fe2 with the spin down. In a supercell, one of the ions belongs
to sublattice Fe1 and three to the sublattice Fe2. The molecular fields
acting on the magnetic moments are
\begin{eqnarray}
H_{2} & = & -\lambda M_{1},\label{eq:HB}\\
H_{1} & = & -\lambda M_{2},\label{eq:HA}\\
\lambda & \equiv & \frac{2J_{1}}{N\mu^{2}}
\end{eqnarray}
 where $N$ is the number of supercells, $\mu=g\mu_{B}$ , $g$ is
the $g$-factor of the Fe$^{3+}$ ion, $\mu_{B}$ is the Bohr magneton,
$M_{1}$ ($M_{2}$) is the magnetization of Fe1(Fe2) sublattice. 
The magnetization,
in its turn, is defined by the molecular field
\begin{eqnarray}
M_{s} & = & Nn_{s}\mu SB_{S}(\frac{\mu S}{k_{B}T}H_{s})
\label{eq:MBs}\\
 & \approx & \frac{C_{s}}{T}H_{s},
\label{eq:M}\end{eqnarray}
 where $s=1,2$, $n_{s}$ is the number of ions in the supercell that
belongs to the sublattice $s$, $n_{2}=3n_{1}=3$, \[
C_{s}=\frac{Nn_{s}\mu^{2}S(S+1)}{3k_{B}}\]
 is the corresponding Curie constant, 
$B_{L}(x)\equiv\left[\left(2L+1\right)/2L\right]\coth
\left[\left(2L+1\right)x/2L\right]-
\left(1/2L\right)\coth x/2L$
is the Brillouin function. The equality (\ref{eq:M}) follows from the 
expansion
$B_{L}(x)\approx\left(L+1\right)x/3L$ , which is valid for small
$x$. Substituting the value of molecular field from Eqs.(\ref{eq:HB})
(\ref{eq:HA})
into Eqs.(\ref{eq:M}), we obtain the system of equations for the
sublattice magnetizations in the absence of the external field
\begin{equation}
\left\{ \begin{array}{ccc}
TM_{1}+C_{1}\lambda M_{2} & = & 0,\\
C_{2}\lambda M_{1}+TM_{2} & = & 0,\end{array}\right.\label{eq:Sys}
\end{equation}
which has trivial solution $M_{1}=M_{2}=0$ above the transition 
temperature
$T>T_{2fe}$. Nonzero values of the magnetizations becomes possible
if the determinant of the coefficients of $M_{1}$ and $M_{2}$ is
zero. This condition yields
\begin{equation}
\left(T_{fe}\right)^{2}=C_{1}C_{2}\left(\frac{2J_{1}}
{N\mu^{2}}\right)^{2},
\label{eq:Tfe}\end{equation}
 and we obtain Eq.(\ref{eq:T2fe}). At lower temperatures $T<T_{fe}$,
the sytem becomes non-linear as the argument of the Brillouin function
grows.

The calculation for PFB3a magnetic ordering is more involved. We have
four sublattices shown on the Fig.\ref{mstr}, Curie constants are
equal $C=N\mu^{2}S(S+1)/3k_{B}$, and we have four equations
\begin{equation}
\left\{ \begin{array}{ccc}
TM_{B,1}= & CH_{B,1}= & -C\lambda\left(M_{A,1}+M_{A,2}\right),\\
TM_{B,2}= & CH_{B,2}= & -C\lambda M_{A,1},\\
TM_{A,1}= & CH_{A,1}= & -C\lambda\left(M_{B,1}+M_{B,2}\right),\\
TM_{A,2}= & CH_{A,2}= & -C\lambda M_{B,1}.\end{array}\right.
\label{eq:Sys3}
\end{equation}
Again, at the transition temperature, the determinant of the coefficients
should vanish. This gives a biquadratic equation\begin{equation}
T^{4}-3\left(C\lambda\right)^{2}T^{2}+\left(C\lambda\right)^{4}=0.
\label{eq:biqu}\end{equation}
 The transition temperature is given by largest positive root of the
Eq.(\ref{eq:biqu}), as it corresponds to the temperature where the
non-trivial solution appears when we approach the transiton from the
paramagnetic side. We thus obtain the Eq. (\ref{eq:T3a}).

\bibliography{pfn}

\begin{thebibliography}{59}%
\makeatletter
\providecommand \@ifxundefined [1]{%
 \@ifx{#1\undefined}
}%
\providecommand \@ifnum [1]{%
 \ifnum #1\expandafter \@firstoftwo
 \else \expandafter \@secondoftwo
 \fi
}%
\providecommand \@ifx [1]{%
 \ifx #1\expandafter \@firstoftwo
 \else \expandafter \@secondoftwo
 \fi
}%
\providecommand \natexlab [1]{#1}%
\providecommand \enquote  [1]{``#1''}%
\providecommand \bibnamefont  [1]{#1}%
\providecommand \bibfnamefont [1]{#1}%
\providecommand \citenamefont [1]{#1}%
\providecommand \href@noop [0]{\@secondoftwo}%
\providecommand \href [0]{\begingroup \@sanitize@url \@href}%
\providecommand \@href[1]{\@@startlink{#1}\@@href}%
\providecommand \@@href[1]{\endgroup#1\@@endlink}%
\providecommand \@sanitize@url [0]{\catcode `\\12\catcode `\$12\catcode
  `\&12\catcode `\#12\catcode `\^12\catcode `\_12\catcode `\%12\relax}%
\providecommand \@@startlink[1]{}%
\providecommand \@@endlink[0]{}%
\providecommand \url  [0]{\begingroup\@sanitize@url \@url }%
\providecommand \@url [1]{\endgroup\@href {#1}{\urlprefix }}%
\providecommand \urlprefix  [0]{URL }%
\providecommand \Eprint [0]{\href }%
\providecommand \doibase [0]{http://dx.doi.org/}%
\providecommand \selectlanguage [0]{\@gobble}%
\providecommand \bibinfo  [0]{\@secondoftwo}%
\providecommand \bibfield  [0]{\@secondoftwo}%
\providecommand \translation [1]{[#1]}%
\providecommand \BibitemOpen [0]{}%
\providecommand \bibitemStop [0]{}%
\providecommand \bibitemNoStop [0]{.\EOS\space}%
\providecommand \EOS [0]{\spacefactor3000\relax}%
\providecommand \BibitemShut  [1]{\csname bibitem#1\endcsname}%
\let\auto@bib@innerbib\@empty
\bibitem [{\citenamefont {Smolenskii}\ and\ \citenamefont
  {Loffe}(1958)}]{Smolenskii58}%
  \BibitemOpen
  \bibfield  {author} {\bibinfo {author} {\bibfnamefont {G.}~\bibnamefont
  {Smolenskii}}\ and\ \bibinfo {author} {\bibfnamefont {V.~A.}\ \bibnamefont
  {Loffe}},\ }\href@noop {} {\bibfield  {journal} {\bibinfo  {journal}
  {Communication No. 71. Grenoble: Colloque International du Magnetisme;}\ }
  (\bibinfo {year} {1958})}\BibitemShut {NoStop}%
\bibitem [{\citenamefont {Bokov}\ \emph {et~al.}(1962)\citenamefont {Bokov},
  \citenamefont {Myl'nikova},\ and\ \citenamefont {Smolenskii}}]{Bokov62}%
  \BibitemOpen
  \bibfield  {author} {\bibinfo {author} {\bibfnamefont {V.}~\bibnamefont
  {Bokov}}, \bibinfo {author} {\bibfnamefont {I.}~\bibnamefont {Myl'nikova}}, \
  and\ \bibinfo {author} {\bibfnamefont {G.~A.}\ \bibnamefont {Smolenskii}},\
  }\href@noop {} {\bibfield  {journal} {\bibinfo  {journal} {Sov. Phys. JETP}\
  }\textbf {\bibinfo {volume} {15}},\ \bibinfo {pages} {447} (\bibinfo {year}
  {1962})}\BibitemShut {NoStop}%
\bibitem [{\citenamefont {Raevski}\ \emph {et~al.}(2009)\citenamefont
  {Raevski}, \citenamefont {Kubrin}, \citenamefont {Raevskaya}, \citenamefont
  {Titov}, \citenamefont {Sarychev}, \citenamefont {Malitskaya}, \citenamefont
  {Zakharchenko},\ and\ \citenamefont {Prosandeev}}]{Raevski09}%
  \BibitemOpen
  \bibfield  {author} {\bibinfo {author} {\bibfnamefont {I.~P.}\ \bibnamefont
  {Raevski}}, \bibinfo {author} {\bibfnamefont {S.~P.}\ \bibnamefont {Kubrin}},
  \bibinfo {author} {\bibfnamefont {S.~I.}\ \bibnamefont {Raevskaya}}, \bibinfo
  {author} {\bibfnamefont {V.~V.}\ \bibnamefont {Titov}}, \bibinfo {author}
  {\bibfnamefont {D.~A.}\ \bibnamefont {Sarychev}}, \bibinfo {author}
  {\bibfnamefont {M.~A.}\ \bibnamefont {Malitskaya}}, \bibinfo {author}
  {\bibfnamefont {I.~N.}\ \bibnamefont {Zakharchenko}}, \ and\ \bibinfo
  {author} {\bibfnamefont {S.~A.}\ \bibnamefont {Prosandeev}},\ }\href
  {\doibase 10.1103/PhysRevB.80.024108} {\bibfield  {journal} {\bibinfo
  {journal} {Phys. Rev. B}\ }\textbf {\bibinfo {volume} {80}},\ \bibinfo
  {pages} {024108} (\bibinfo {year} {2009})}\BibitemShut {NoStop}%
\bibitem [{\citenamefont {Rotaru}\ \emph {et~al.}(2009)\citenamefont {Rotaru},
  \citenamefont {Roessli}, \citenamefont {Amato}, \citenamefont {Gvasaliya},
  \citenamefont {Mudry}, \citenamefont {Lushnikov},\ and\ \citenamefont
  {Shaplygina}}]{Rotaru09}%
  \BibitemOpen
  \bibfield  {author} {\bibinfo {author} {\bibfnamefont {G.~M.}\ \bibnamefont
  {Rotaru}}, \bibinfo {author} {\bibfnamefont {B.}~\bibnamefont {Roessli}},
  \bibinfo {author} {\bibfnamefont {A.}~\bibnamefont {Amato}}, \bibinfo
  {author} {\bibfnamefont {S.~N.}\ \bibnamefont {Gvasaliya}}, \bibinfo {author}
  {\bibfnamefont {C.}~\bibnamefont {Mudry}}, \bibinfo {author} {\bibfnamefont
  {S.~G.}\ \bibnamefont {Lushnikov}}, \ and\ \bibinfo {author} {\bibfnamefont
  {T.~A.}\ \bibnamefont {Shaplygina}},\ }\href {\doibase
  10.1103/PhysRevB.79.184430} {\bibfield  {journal} {\bibinfo  {journal} {Phys.
  Rev. B}\ }\textbf {\bibinfo {volume} {79}},\ \bibinfo {pages} {184430}
  (\bibinfo {year} {2009})}\BibitemShut {NoStop}%
\bibitem [{\citenamefont {Kleemann}\ \emph {et~al.}(2010)\citenamefont
  {Kleemann}, \citenamefont {Shvartsman}, \citenamefont {Borisov},\ and\
  \citenamefont {Kania}}]{Kleemann10}%
  \BibitemOpen
  \bibfield  {author} {\bibinfo {author} {\bibfnamefont {W.}~\bibnamefont
  {Kleemann}}, \bibinfo {author} {\bibfnamefont {V.~V.}\ \bibnamefont
  {Shvartsman}}, \bibinfo {author} {\bibfnamefont {P.}~\bibnamefont {Borisov}},
  \ and\ \bibinfo {author} {\bibfnamefont {A.}~\bibnamefont {Kania}},\ }\href
  {\doibase 10.1103/PhysRevLett.105.257202} {\bibfield  {journal} {\bibinfo
  {journal} {Phys. Rev. Lett.}\ }\textbf {\bibinfo {volume} {105}},\ \bibinfo
  {pages} {257202} (\bibinfo {year} {2010})}\BibitemShut {NoStop}%
\bibitem [{\citenamefont {Laguta}\ \emph {et~al.}(2010)\citenamefont {Laguta},
  \citenamefont {Rosa}, \citenamefont {Jastrabik}, \citenamefont {Blinc},
  \citenamefont {Cevc}, \citenamefont {Zalar}, \citenamefont {Remskar},
  \citenamefont {Raevskaya},\ and\ \citenamefont {Raevski}}]{Laguta10}%
  \BibitemOpen
  \bibfield  {author} {\bibinfo {author} {\bibfnamefont {V.}~\bibnamefont
  {Laguta}}, \bibinfo {author} {\bibfnamefont {J.}~\bibnamefont {Rosa}},
  \bibinfo {author} {\bibfnamefont {L.}~\bibnamefont {Jastrabik}}, \bibinfo
  {author} {\bibfnamefont {R.}~\bibnamefont {Blinc}}, \bibinfo {author}
  {\bibfnamefont {P.}~\bibnamefont {Cevc}}, \bibinfo {author} {\bibfnamefont
  {B.}~\bibnamefont {Zalar}}, \bibinfo {author} {\bibfnamefont
  {M.}~\bibnamefont {Remskar}}, \bibinfo {author} {\bibfnamefont
  {S.}~\bibnamefont {Raevskaya}}, \ and\ \bibinfo {author} {\bibfnamefont
  {I.}~\bibnamefont {Raevski}},\ }\href@noop {} {\bibfield  {journal} {\bibinfo
   {journal} {Mater. Res. Bull.}\ }\textbf {\bibinfo {volume} {45}},\ \bibinfo
  {pages} {1720} (\bibinfo {year} {2010})}\BibitemShut {NoStop}%
\bibitem [{\citenamefont {Raevski}\ \emph {et~al.}(2012)\citenamefont
  {Raevski}, \citenamefont {Kubrin}, \citenamefont {Raevskaya}, \citenamefont
  {Sarychev}, \citenamefont {Prosandeev},\ and\ \citenamefont
  {Malitskaya}}]{Raevski12}%
  \BibitemOpen
  \bibfield  {author} {\bibinfo {author} {\bibfnamefont {I.~P.}\ \bibnamefont
  {Raevski}}, \bibinfo {author} {\bibfnamefont {S.~P.}\ \bibnamefont {Kubrin}},
  \bibinfo {author} {\bibfnamefont {S.~I.}\ \bibnamefont {Raevskaya}}, \bibinfo
  {author} {\bibfnamefont {D.~A.}\ \bibnamefont {Sarychev}}, \bibinfo {author}
  {\bibfnamefont {S.~A.}\ \bibnamefont {Prosandeev}}, \ and\ \bibinfo {author}
  {\bibfnamefont {M.~A.}\ \bibnamefont {Malitskaya}},\ }\href {\doibase
  10.1103/PhysRevB.85.224412} {\bibfield  {journal} {\bibinfo  {journal} {Phys.
  Rev. B}\ }\textbf {\bibinfo {volume} {85}},\ \bibinfo {pages} {224412}
  (\bibinfo {year} {2012})}\BibitemShut {NoStop}%
\bibitem [{\citenamefont {Laguta}\ \emph {et~al.}(2013)\citenamefont {Laguta},
  \citenamefont {Glinchuk}, \citenamefont {Mary\v{s}ko}, \citenamefont
  {Kuzian}, \citenamefont {Prosandeev}, \citenamefont {Raevskaya},
  \citenamefont {Smotrakov}, \citenamefont {Eremkin},\ and\ \citenamefont
  {Raevski}}]{Laguta13}%
  \BibitemOpen
  \bibfield  {author} {\bibinfo {author} {\bibfnamefont {V.~V.}\ \bibnamefont
  {Laguta}}, \bibinfo {author} {\bibfnamefont {M.~D.}\ \bibnamefont
  {Glinchuk}}, \bibinfo {author} {\bibfnamefont {M.}~\bibnamefont
  {Mary\v{s}ko}}, \bibinfo {author} {\bibfnamefont {R.~O.}\ \bibnamefont
  {Kuzian}}, \bibinfo {author} {\bibfnamefont {S.~A.}\ \bibnamefont
  {Prosandeev}}, \bibinfo {author} {\bibfnamefont {S.~I.}\ \bibnamefont
  {Raevskaya}}, \bibinfo {author} {\bibfnamefont {V.~G.}\ \bibnamefont
  {Smotrakov}}, \bibinfo {author} {\bibfnamefont {V.~V.}\ \bibnamefont
  {Eremkin}}, \ and\ \bibinfo {author} {\bibfnamefont {I.~P.}\ \bibnamefont
  {Raevski}},\ }\href {\doibase 10.1103/PhysRevB.87.064403} {\bibfield
  {journal} {\bibinfo  {journal} {Phys. Rev. B}\ }\textbf {\bibinfo {volume}
  {87}},\ \bibinfo {pages} {064403} (\bibinfo {year} {2013})}\BibitemShut
  {NoStop}%
\bibitem [{\citenamefont {Kubrin}(2009)}]{Kubrin}%
  \BibitemOpen
  \bibfield  {author} {\bibinfo {author} {\bibfnamefont {S.}~\bibnamefont
  {Kubrin}},\ }\emph {\bibinfo {title} {Magnetic phase transitions in ternary
  iron oxides with perovskite structure studied using M\"{o}ssbauer
  spectroscopy}},\ \href@noop {} {Ph.D. thesis},\ \bibinfo  {school} {Southern
  Federal University}, \bibinfo {address} {Rostov on Don 344090, Russia}
  (\bibinfo {year} {2009})\BibitemShut {NoStop}%
\bibitem [{\citenamefont {Treves}(1965)}]{Treves65}%
  \BibitemOpen
  \bibfield  {author} {\bibinfo {author} {\bibfnamefont {D.}~\bibnamefont
  {Treves}},\ }\href@noop {} {\bibfield  {journal} {\bibinfo  {journal} {J.
  Appl. Phys.}\ }\textbf {\bibinfo {volume} {36}},\ \bibinfo {pages} {1033}
  (\bibinfo {year} {1965})}\BibitemShut {NoStop}%
\bibitem [{\citenamefont {Eibsch\"utz}\ \emph {et~al.}(1967)\citenamefont
  {Eibsch\"utz}, \citenamefont {Shtrikman},\ and\ \citenamefont
  {Treves}}]{Eibschuetz67}%
  \BibitemOpen
  \bibfield  {author} {\bibinfo {author} {\bibfnamefont {M.}~\bibnamefont
  {Eibsch\"utz}}, \bibinfo {author} {\bibfnamefont {S.}~\bibnamefont
  {Shtrikman}}, \ and\ \bibinfo {author} {\bibfnamefont {D.}~\bibnamefont
  {Treves}},\ }\href {\doibase 10.1103/PhysRev.156.562} {\bibfield  {journal}
  {\bibinfo  {journal} {Phys. Rev.}\ }\textbf {\bibinfo {volume} {156}},\
  \bibinfo {pages} {562} (\bibinfo {year} {1967})}\BibitemShut {NoStop}%
\bibitem [{\citenamefont {Gabbasova}\ \emph {et~al.}(1991)\citenamefont
  {Gabbasova}, \citenamefont {Kuz'min}, \citenamefont {Zvezdin}, \citenamefont
  {Dubenko}, \citenamefont {Murashov}, \citenamefont {Rakov},\ and\
  \citenamefont {Krynetsky}}]{Gabbasova91}%
  \BibitemOpen
  \bibfield  {author} {\bibinfo {author} {\bibfnamefont {Z.~V.}\ \bibnamefont
  {Gabbasova}}, \bibinfo {author} {\bibfnamefont {M.~D.}\ \bibnamefont
  {Kuz'min}}, \bibinfo {author} {\bibfnamefont {A.~K.}\ \bibnamefont
  {Zvezdin}}, \bibinfo {author} {\bibfnamefont {I.~S.}\ \bibnamefont
  {Dubenko}}, \bibinfo {author} {\bibfnamefont {V.~A.}\ \bibnamefont
  {Murashov}}, \bibinfo {author} {\bibfnamefont {D.~N.}\ \bibnamefont {Rakov}},
  \ and\ \bibinfo {author} {\bibfnamefont {I.~B.}\ \bibnamefont {Krynetsky}},\
  }\href@noop {} {\bibfield  {journal} {\bibinfo  {journal} {Phys.\ Lett.\ A}\
  }\textbf {\bibinfo {volume} {158}},\ \bibinfo {pages} {491} (\bibinfo {year}
  {1991})}\BibitemShut {NoStop}%
\bibitem [{\citenamefont {Gorodetsky}(1969)}]{Gorodetsky69}%
  \BibitemOpen
  \bibfield  {author} {\bibinfo {author} {\bibfnamefont {G.}~\bibnamefont
  {Gorodetsky}},\ }\href@noop {} {\bibfield  {journal} {\bibinfo  {journal}
  {J.\ Phys.\ Chem.\ Solids}\ }\textbf {\bibinfo {volume} {30}},\ \bibinfo
  {pages} {1745} (\bibinfo {year} {1969})}\BibitemShut {NoStop}%
\bibitem [{\citenamefont {Shapiro}\ \emph {et~al.}(1974)\citenamefont
  {Shapiro}, \citenamefont {Axe},\ and\ \citenamefont {Remeika}}]{Shapiro74}%
  \BibitemOpen
  \bibfield  {author} {\bibinfo {author} {\bibfnamefont {S.~M.}\ \bibnamefont
  {Shapiro}}, \bibinfo {author} {\bibfnamefont {J.~D.}\ \bibnamefont {Axe}}, \
  and\ \bibinfo {author} {\bibfnamefont {J.~P.}\ \bibnamefont {Remeika}},\
  }\href {\doibase 10.1103/PhysRevB.10.2014} {\bibfield  {journal} {\bibinfo
  {journal} {Phys. Rev. B}\ }\textbf {\bibinfo {volume} {10}},\ \bibinfo
  {pages} {2014} (\bibinfo {year} {1974})}\BibitemShut {NoStop}%
\bibitem [{\citenamefont {Gukasov}\ \emph {et~al.}(1997)\citenamefont
  {Gukasov}, \citenamefont {Steigenberger}, \citenamefont {Barilo},\ and\
  \citenamefont {Guretskii}}]{Gukasov97}%
  \BibitemOpen
  \bibfield  {author} {\bibinfo {author} {\bibfnamefont {A.}~\bibnamefont
  {Gukasov}}, \bibinfo {author} {\bibfnamefont {U.}~\bibnamefont
  {Steigenberger}}, \bibinfo {author} {\bibfnamefont {S.}~\bibnamefont
  {Barilo}}, \ and\ \bibinfo {author} {\bibfnamefont {S.}~\bibnamefont
  {Guretskii}},\ }\href {\doibase
  http://dx.doi.org/10.1016/S0921-4526(96)01156-8} {\bibfield  {journal}
  {\bibinfo  {journal} {Physica B: Condensed Matter}\ }\textbf {\bibinfo
  {volume} {234–236}},\ \bibinfo {pages} {760 } (\bibinfo {year} {1997})},\
  \bibinfo {note} {<ce:title>Proceedings of the First European Conference on
  Neutron Scattering</ce:title>}\BibitemShut {NoStop}%
\bibitem [{\citenamefont {Delaire}\ \emph {et~al.}(2012)\citenamefont
  {Delaire}, \citenamefont {Stone}, \citenamefont {Ma}, \citenamefont {Huq},
  \citenamefont {Gout}, \citenamefont {Brown}, \citenamefont {Wang},\ and\
  \citenamefont {Ren}}]{Delaire12}%
  \BibitemOpen
  \bibfield  {author} {\bibinfo {author} {\bibfnamefont {O.}~\bibnamefont
  {Delaire}}, \bibinfo {author} {\bibfnamefont {M.~B.}\ \bibnamefont {Stone}},
  \bibinfo {author} {\bibfnamefont {J.}~\bibnamefont {Ma}}, \bibinfo {author}
  {\bibfnamefont {A.}~\bibnamefont {Huq}}, \bibinfo {author} {\bibfnamefont
  {D.}~\bibnamefont {Gout}}, \bibinfo {author} {\bibfnamefont {C.}~\bibnamefont
  {Brown}}, \bibinfo {author} {\bibfnamefont {K.~F.}\ \bibnamefont {Wang}}, \
  and\ \bibinfo {author} {\bibfnamefont {Z.~F.}\ \bibnamefont {Ren}},\ }\href
  {\doibase 10.1103/PhysRevB.85.064405} {\bibfield  {journal} {\bibinfo
  {journal} {Phys. Rev. B}\ }\textbf {\bibinfo {volume} {85}},\ \bibinfo
  {pages} {064405} (\bibinfo {year} {2012})}\BibitemShut {NoStop}%
\bibitem [{\citenamefont {McQueeney}\ \emph {et~al.}(2008)\citenamefont
  {McQueeney}, \citenamefont {Yan}, \citenamefont {Chang},\ and\ \citenamefont
  {Ma}}]{McQueeney08}%
  \BibitemOpen
  \bibfield  {author} {\bibinfo {author} {\bibfnamefont {R.~J.}\ \bibnamefont
  {McQueeney}}, \bibinfo {author} {\bibfnamefont {J.-Q.}\ \bibnamefont {Yan}},
  \bibinfo {author} {\bibfnamefont {S.}~\bibnamefont {Chang}}, \ and\ \bibinfo
  {author} {\bibfnamefont {J.}~\bibnamefont {Ma}},\ }\href {\doibase
  10.1103/PhysRevB.78.184417} {\bibfield  {journal} {\bibinfo  {journal} {Phys.
  Rev. B}\ }\textbf {\bibinfo {volume} {78}},\ \bibinfo {pages} {184417}
  (\bibinfo {year} {2008})}\BibitemShut {NoStop}%
\bibitem [{ort()}]{orthofe}%
  \BibitemOpen
  \href@noop {} {}\bibinfo {note} {The $J_{1},J_{2}$ values for the
  orthoferrites YFeO$_{3}$ and LuFeO$_{3}$ were obtained in
  Ref.\onlinecite{Gorodetsky69} from the high temperature analysis of the
  paramagnetic susceptibility. In our notations, they are
  $J_{1}/k_{B}=54.4\pm2$~K, $J_{2}/k_{B}=2.4\pm0.4$~K, for YFeO$_{3}$, and
  $J_{1}/k_{B}=48.4\pm2$~K, $J_{2}/k_{B}=2\pm0.2$~K for LuFeO$_{3}$. Note that
  the author of the Ref.\onlinecite{Gorodetsky69} use another definition of the
  exchange, and we have multiplied his values by the factor (-2) for the
  comparison with our values. The inelastic neutron scattering
  studies\cite{Shapiro74,Gukasov97,Delaire12,McQueeney08} shows that $J_1,J_2$
  have almost the same values for all RFeO$_3$ family, the small variations
  being due to the Fe-O-Fe angle variations.\cite{Boekema72}}\BibitemShut
  {NoStop}%
\bibitem [{\citenamefont {D'Ariano}\ and\ \citenamefont
  {Borsa}(1982)}]{DAriano82}%
  \BibitemOpen
  \bibfield  {author} {\bibinfo {author} {\bibfnamefont {G.}~\bibnamefont
  {D'Ariano}}\ and\ \bibinfo {author} {\bibfnamefont {F.}~\bibnamefont
  {Borsa}},\ }\href {\doibase 10.1103/PhysRevB.26.6215} {\bibfield  {journal}
  {\bibinfo  {journal} {Phys. Rev. B}\ }\textbf {\bibinfo {volume} {26}},\
  \bibinfo {pages} {6215} (\bibinfo {year} {1982})}\BibitemShut {NoStop}%
\bibitem [{\citenamefont {Breed}\ \emph {et~al.}(1970)\citenamefont {Breed},
  \citenamefont {Gilijamse}, \citenamefont {Sterkenburg},\ and\ \citenamefont
  {Miedema}}]{Breed70}%
  \BibitemOpen
  \bibfield  {author} {\bibinfo {author} {\bibfnamefont {D.~J.}\ \bibnamefont
  {Breed}}, \bibinfo {author} {\bibfnamefont {K.}~\bibnamefont {Gilijamse}},
  \bibinfo {author} {\bibfnamefont {J.~W.~E.}\ \bibnamefont {Sterkenburg}}, \
  and\ \bibinfo {author} {\bibfnamefont {A.~R.}\ \bibnamefont {Miedema}},\
  }\href {http://dx.doi.org.sci-hub.org/10.1063/1.1658906} {\bibfield
  {journal} {\bibinfo  {journal} {J. Appl. Phys.}\ }\textbf {\bibinfo {volume}
  {41}},\ \bibinfo {pages} {1267} (\bibinfo {year} {1970})}\BibitemShut
  {NoStop}%
\bibitem [{\citenamefont {Stinchcombe}(1979)}]{Stinchcombe79}%
  \BibitemOpen
  \bibfield  {author} {\bibinfo {author} {\bibfnamefont {R.~B.}\ \bibnamefont
  {Stinchcombe}},\ }\href {http://stacks.iop.org/0022-3719/12/i=21/a=020}
  {\bibfield  {journal} {\bibinfo  {journal} {Journal of Physics C: Solid State
  Physics}\ }\textbf {\bibinfo {volume} {12}},\ \bibinfo {pages} {4533}
  (\bibinfo {year} {1979})}\BibitemShut {NoStop}%
\bibitem [{\citenamefont {Kumar}\ \emph {et~al.}(1981)\citenamefont {Kumar},
  \citenamefont {Pandey},\ and\ \citenamefont {Barma}}]{Kumar81}%
  \BibitemOpen
  \bibfield  {author} {\bibinfo {author} {\bibfnamefont {D.}~\bibnamefont
  {Kumar}}, \bibinfo {author} {\bibfnamefont {R.~B.}\ \bibnamefont {Pandey}}, \
  and\ \bibinfo {author} {\bibfnamefont {M.}~\bibnamefont {Barma}},\ }\href
  {\doibase 10.1103/PhysRevB.23.2269} {\bibfield  {journal} {\bibinfo
  {journal} {Phys. Rev. B}\ }\textbf {\bibinfo {volume} {23}},\ \bibinfo
  {pages} {2269} (\bibinfo {year} {1981})}\BibitemShut {NoStop}%
\bibitem [{\citenamefont {Battle}\ \emph
  {et~al.}(1995{\natexlab{a}})\citenamefont {Battle}, \citenamefont {Gibb},
  \citenamefont {Herod}, \citenamefont {Kim},\ and\ \citenamefont
  {Munns}}]{Battle95}%
  \BibitemOpen
  \bibfield  {author} {\bibinfo {author} {\bibfnamefont {P.~D.}\ \bibnamefont
  {Battle}}, \bibinfo {author} {\bibfnamefont {T.}~\bibnamefont {Gibb}},
  \bibinfo {author} {\bibfnamefont {A.}~\bibnamefont {Herod}}, \bibinfo
  {author} {\bibfnamefont {S.-H.}\ \bibnamefont {Kim}}, \ and\ \bibinfo
  {author} {\bibfnamefont {P.}~\bibnamefont {Munns}},\ }\href@noop {}
  {\bibfield  {journal} {\bibinfo  {journal} {J. Mater. Chem.}\ }\textbf
  {\bibinfo {volume} {5}},\ \bibinfo {pages} {865} (\bibinfo {year}
  {1995}{\natexlab{a}})}\BibitemShut {NoStop}%
\bibitem [{\citenamefont {Battle}\ \emph
  {et~al.}(1995{\natexlab{b}})\citenamefont {Battle}, \citenamefont {Gibb},
  \citenamefont {Herod},\ and\ \citenamefont {Hodges}}]{Battle95a}%
  \BibitemOpen
  \bibfield  {author} {\bibinfo {author} {\bibfnamefont {P.~D.}\ \bibnamefont
  {Battle}}, \bibinfo {author} {\bibfnamefont {T.}~\bibnamefont {Gibb}},
  \bibinfo {author} {\bibfnamefont {A.}~\bibnamefont {Herod}}, \ and\ \bibinfo
  {author} {\bibfnamefont {J.}~\bibnamefont {Hodges}},\ }\href@noop {}
  {\bibfield  {journal} {\bibinfo  {journal} {J. Mater. Chem.}\ }\textbf
  {\bibinfo {volume} {5}},\ \bibinfo {pages} {75} (\bibinfo {year}
  {1995}{\natexlab{b}})}\BibitemShut {NoStop}%
\bibitem [{\citenamefont {Blinc}\ \emph {et~al.}(2008)\citenamefont {Blinc},
  \citenamefont {Laguta}, \citenamefont {Zalar}, \citenamefont
  {Zupan\u{c}i\u{c}},\ and\ \citenamefont {Itoh}}]{Blinc08}%
  \BibitemOpen
  \bibfield  {author} {\bibinfo {author} {\bibfnamefont {R.}~\bibnamefont
  {Blinc}}, \bibinfo {author} {\bibfnamefont {V.~V.}\ \bibnamefont {Laguta}},
  \bibinfo {author} {\bibfnamefont {B.}~\bibnamefont {Zalar}}, \bibinfo
  {author} {\bibfnamefont {B.}~\bibnamefont {Zupan\u{c}i\u{c}}}, \ and\
  \bibinfo {author} {\bibfnamefont {M.}~\bibnamefont {Itoh}},\ }\href {\doibase
  http://dx.doi.org/10.1063/1.2957077} {\bibfield  {journal} {\bibinfo
  {journal} {Journal of Applied Physics}\ }\textbf {\bibinfo {volume} {104}},\
  \bibinfo {pages} {084105} (\bibinfo {year} {2008})}\BibitemShut {NoStop}%
\bibitem [{\citenamefont {Gu}\ \emph {et~al.}(1999)\citenamefont {Gu},
  \citenamefont {Gui}, \citenamefont {Liu},\ and\ \citenamefont
  {Zhang}}]{Gu99}%
  \BibitemOpen
  \bibfield  {author} {\bibinfo {author} {\bibfnamefont {B.-L.}\ \bibnamefont
  {Gu}}, \bibinfo {author} {\bibfnamefont {H.}~\bibnamefont {Gui}}, \bibinfo
  {author} {\bibfnamefont {Z.-R.}\ \bibnamefont {Liu}}, \ and\ \bibinfo
  {author} {\bibfnamefont {X.-W.}\ \bibnamefont {Zhang}},\ }\href {\doibase
  http://dx.doi.org/10.1063/1.369558} {\bibfield  {journal} {\bibinfo
  {journal} {Journal of Applied Physics}\ }\textbf {\bibinfo {volume} {85}},\
  \bibinfo {pages} {2408} (\bibinfo {year} {1999})}\BibitemShut {NoStop}%
\bibitem [{rta()}]{rtab}%
  \BibitemOpen
  \href@noop {} {}\bibinfo {note} {Note that, in the
  Ref.\onlinecite{Raevski12}, the rows 2 and 4 of the Table.II are interchanged
  compared with the Fig.3. Actually, the PFB2 configuration has the energy
  close to the PFB5 one.}\BibitemShut {Stop}%
\bibitem [{\citenamefont {Sanchez}\ \emph {et~al.}(2013)\citenamefont
  {Sanchez}, \citenamefont {Ortega}, \citenamefont {Kumar}, \citenamefont
  {Sreenivasulu}, \citenamefont {Katiyar}, \citenamefont {Scott}, \citenamefont
  {Evans}, \citenamefont {Arredondo-Arechavala}, \citenamefont {Schilling},\
  and\ \citenamefont {Gregg}}]{Sanchez13}%
  \BibitemOpen
  \bibfield  {author} {\bibinfo {author} {\bibfnamefont {D.~A.}\ \bibnamefont
  {Sanchez}}, \bibinfo {author} {\bibfnamefont {N.}~\bibnamefont {Ortega}},
  \bibinfo {author} {\bibfnamefont {A.}~\bibnamefont {Kumar}}, \bibinfo
  {author} {\bibfnamefont {G.}~\bibnamefont {Sreenivasulu}}, \bibinfo {author}
  {\bibfnamefont {R.~S.}\ \bibnamefont {Katiyar}}, \bibinfo {author}
  {\bibfnamefont {J.~F.}\ \bibnamefont {Scott}}, \bibinfo {author}
  {\bibfnamefont {D.~M.}\ \bibnamefont {Evans}}, \bibinfo {author}
  {\bibfnamefont {M.}~\bibnamefont {Arredondo-Arechavala}}, \bibinfo {author}
  {\bibfnamefont {A.}~\bibnamefont {Schilling}}, \ and\ \bibinfo {author}
  {\bibfnamefont {J.~M.}\ \bibnamefont {Gregg}},\ }\href {\doibase
  10.1063/1.4790317} {\bibfield  {journal} {\bibinfo  {journal} {Journal of
  Applied Physics}\ }\textbf {\bibinfo {volume} {113}},\ \bibinfo {pages}
  {074105} (\bibinfo {year} {2013})}\BibitemShut {NoStop}%
\bibitem [{\citenamefont {Sanchez}\ \emph {et~al.}(2011)\citenamefont
  {Sanchez}, \citenamefont {Ortega}, \citenamefont {Kumar}, \citenamefont
  {Roque-Malherbe}, \citenamefont {Polanco}, \citenamefont {Scott},\ and\
  \citenamefont {Katiyar}}]{Sanchez11}%
  \BibitemOpen
  \bibfield  {author} {\bibinfo {author} {\bibfnamefont {D.~A.}\ \bibnamefont
  {Sanchez}}, \bibinfo {author} {\bibfnamefont {N.}~\bibnamefont {Ortega}},
  \bibinfo {author} {\bibfnamefont {A.}~\bibnamefont {Kumar}}, \bibinfo
  {author} {\bibfnamefont {R.}~\bibnamefont {Roque-Malherbe}}, \bibinfo
  {author} {\bibfnamefont {R.}~\bibnamefont {Polanco}}, \bibinfo {author}
  {\bibfnamefont {J.~F.}\ \bibnamefont {Scott}}, \ and\ \bibinfo {author}
  {\bibfnamefont {R.~S.}\ \bibnamefont {Katiyar}},\ }\href {\doibase
  10.1063/1.3670361} {\bibfield  {journal} {\bibinfo  {journal} {AIP Advances}\
  }\textbf {\bibinfo {volume} {1}},\ \bibinfo {pages} {042169} (\bibinfo {year}
  {2011})}\BibitemShut {NoStop}%
\bibitem [{\citenamefont {Evans}\ \emph {et~al.}(2013)\citenamefont {Evans},
  \citenamefont {Schilling}, \citenamefont {Kumar}, \citenamefont {Sanchez},
  \citenamefont {Ortega}, \citenamefont {Arredondo}, \citenamefont {Katiyar},
  \citenamefont {Gregg},\ and\ \citenamefont {Scott}}]{Evans13}%
  \BibitemOpen
  \bibfield  {author} {\bibinfo {author} {\bibfnamefont {D.}~\bibnamefont
  {Evans}}, \bibinfo {author} {\bibfnamefont {A.}~\bibnamefont {Schilling}},
  \bibinfo {author} {\bibfnamefont {A.}~\bibnamefont {Kumar}}, \bibinfo
  {author} {\bibfnamefont {D.}~\bibnamefont {Sanchez}}, \bibinfo {author}
  {\bibfnamefont {N.}~\bibnamefont {Ortega}}, \bibinfo {author} {\bibfnamefont
  {M.}~\bibnamefont {Arredondo}}, \bibinfo {author} {\bibfnamefont
  {R.}~\bibnamefont {Katiyar}}, \bibinfo {author} {\bibfnamefont
  {J.}~\bibnamefont {Gregg}}, \ and\ \bibinfo {author} {\bibfnamefont
  {J.}~\bibnamefont {Scott}},\ }\href {\doibase 10.1038/ncomms2548} {\bibfield
  {journal} {\bibinfo  {journal} {Nat. Commun.}\ }\textbf {\bibinfo {volume}
  {4}},\ \bibinfo {pages} {1534} (\bibinfo {year} {2013})}\BibitemShut
  {NoStop}%
\bibitem [{\citenamefont {Kumar}\ \emph {et~al.}(2009)\citenamefont {Kumar},
  \citenamefont {Sharma}, \citenamefont {Katiyar}, \citenamefont {Pirc},
  \citenamefont {Blinc},\ and\ \citenamefont {Scott}}]{Kumar09}%
  \BibitemOpen
  \bibfield  {author} {\bibinfo {author} {\bibfnamefont {A.}~\bibnamefont
  {Kumar}}, \bibinfo {author} {\bibfnamefont {G.~L.}\ \bibnamefont {Sharma}},
  \bibinfo {author} {\bibfnamefont {R.~S.}\ \bibnamefont {Katiyar}}, \bibinfo
  {author} {\bibfnamefont {R.}~\bibnamefont {Pirc}}, \bibinfo {author}
  {\bibfnamefont {R.}~\bibnamefont {Blinc}}, \ and\ \bibinfo {author}
  {\bibfnamefont {J.~F.}\ \bibnamefont {Scott}},\ }\href
  {http://stacks.iop.org/0953-8984/21/i=38/a=382204} {\bibfield  {journal}
  {\bibinfo  {journal} {Journal of Physics: Condensed Matter}\ }\textbf
  {\bibinfo {volume} {21}},\ \bibinfo {pages} {382204} (\bibinfo {year}
  {2009})}\BibitemShut {NoStop}%
\bibitem [{\citenamefont {Lampis}\ \emph {et~al.}(2004)\citenamefont {Lampis},
  \citenamefont {Franchini}, \citenamefont {Satta}, \citenamefont
  {Geddo-Lehmann},\ and\ \citenamefont {Massidda}}]{Lampis04}%
  \BibitemOpen
  \bibfield  {author} {\bibinfo {author} {\bibfnamefont {N.}~\bibnamefont
  {Lampis}}, \bibinfo {author} {\bibfnamefont {C.}~\bibnamefont {Franchini}},
  \bibinfo {author} {\bibfnamefont {G.}~\bibnamefont {Satta}}, \bibinfo
  {author} {\bibfnamefont {A.}~\bibnamefont {Geddo-Lehmann}}, \ and\ \bibinfo
  {author} {\bibfnamefont {S.}~\bibnamefont {Massidda}},\ }\href {\doibase
  10.1103/PhysRevB.69.064412} {\bibfield  {journal} {\bibinfo  {journal} {Phys.
  Rev. B}\ }\textbf {\bibinfo {volume} {69}},\ \bibinfo {pages} {064412}
  (\bibinfo {year} {2004})}\BibitemShut {NoStop}%
\bibitem [{FPL()}]{FPLO}%
  \BibitemOpen
  \href {http://www.FPLO.de.} {}\bibinfo {note} {FPLO-9.00-34 [improved version
  of the original FPLO code by K.\ Koepernik and H.\ Eschrig, Phys.\ Rev.\ B
  \textbf{59}, 1743 (1999)];http://www.FPLO.de}\BibitemShut {NoStop}%
\bibitem [{\citenamefont {Perdew}\ and\ \citenamefont {Wang}(1992)}]{PW}%
  \BibitemOpen
  \bibfield  {author} {\bibinfo {author} {\bibfnamefont {J.~P.}\ \bibnamefont
  {Perdew}}\ and\ \bibinfo {author} {\bibfnamefont {Y.}~\bibnamefont {Wang}},\
  }\href {\doibase 10.1103/PhysRevB.45.13244} {\bibfield  {journal} {\bibinfo
  {journal} {Phys. Rev. B}\ }\textbf {\bibinfo {volume} {45}},\ \bibinfo
  {pages} {13244} (\bibinfo {year} {1992})}\BibitemShut {NoStop}%
\bibitem [{\citenamefont {Eschrig}\ \emph {et~al.}(2003)\citenamefont
  {Eschrig}, \citenamefont {Koepernik},\ and\ \citenamefont
  {Chaplygin}}]{Eschrig03}%
  \BibitemOpen
  \bibfield  {author} {\bibinfo {author} {\bibfnamefont {H.}~\bibnamefont
  {Eschrig}}, \bibinfo {author} {\bibfnamefont {K.}~\bibnamefont {Koepernik}},
  \ and\ \bibinfo {author} {\bibfnamefont {I.}~\bibnamefont {Chaplygin}},\
  }\href@noop {} {\bibfield  {journal} {\bibinfo  {journal} {J.\ Solid State
  Chem.}\ }\textbf {\bibinfo {volume} {176}},\ \bibinfo {pages} {482} (\bibinfo
  {year} {2003})}\BibitemShut {NoStop}%
\bibitem [{\citenamefont {Czyzyk}\ and\ \citenamefont
  {Sawatzky}(1994)}]{Czyzyk94}%
  \BibitemOpen
  \bibfield  {author} {\bibinfo {author} {\bibfnamefont {M.~T.}\ \bibnamefont
  {Czyzyk}}\ and\ \bibinfo {author} {\bibfnamefont {G.~A.}\ \bibnamefont
  {Sawatzky}},\ }\href {\doibase 10.1103/PhysRevB.49.14211} {\bibfield
  {journal} {\bibinfo  {journal} {Phys. Rev. B}\ }\textbf {\bibinfo {volume}
  {49}},\ \bibinfo {pages} {14211} (\bibinfo {year} {1994})}\BibitemShut
  {NoStop}%
\bibitem [{\citenamefont {Tanabe}\ and\ \citenamefont
  {Sugano}(1954)}]{Tanabe54}%
  \BibitemOpen
  \bibfield  {author} {\bibinfo {author} {\bibfnamefont {Y.}~\bibnamefont
  {Tanabe}}\ and\ \bibinfo {author} {\bibfnamefont {S.}~\bibnamefont
  {Sugano}},\ }\href@noop {} {\bibfield  {journal} {\bibinfo  {journal} {J.
  Phys. Soc. Jpn.}\ }\textbf {\bibinfo {volume} {9}},\ \bibinfo {pages} {766}
  (\bibinfo {year} {1954})}\BibitemShut {NoStop}%
\bibitem [{rae()}]{raev2}%
  \BibitemOpen
  \href@noop {} {}\bibinfo {note} {Courtesy of the authors of the
  Ref.\onlinecite{Raevski12}.}\BibitemShut {Stop}%
\bibitem [{\citenamefont {{Kuzian}}\ \emph {et~al.}(2013)\citenamefont
  {{Kuzian}}, \citenamefont {{Laguta}},\ and\ \citenamefont
  {{Richter}}}]{Kuzian13arX}%
  \BibitemOpen
  \bibfield  {author} {\bibinfo {author} {\bibfnamefont {R.~O.}\ \bibnamefont
  {{Kuzian}}}, \bibinfo {author} {\bibfnamefont {V.~V.}\ \bibnamefont
  {{Laguta}}}, \ and\ \bibinfo {author} {\bibfnamefont {J.}~\bibnamefont
  {{Richter}}},\ }\href@noop {} {\bibfield  {journal} {\bibinfo  {journal}
  {ArXiv e-prints}\ } (\bibinfo {year} {2013})},\ \Eprint
  {http://arxiv.org/abs/1310.8079} {arXiv:1310.8079 [cond-mat.mtrl-sci]}
  \BibitemShut {NoStop}%
\bibitem [{\citenamefont {Anderson}(1963)}]{Anderson63}%
  \BibitemOpen
  \bibfield  {author} {\bibinfo {author} {\bibfnamefont {P.~W.}\ \bibnamefont
  {Anderson}}\ }(\bibinfo  {publisher} {Academic Press},\ \bibinfo {year}
  {1963})\ pp.\ \bibinfo {pages} {99--214}\BibitemShut {NoStop}%
\bibitem [{\citenamefont {Larson}\ \emph {et~al.}(1988)\citenamefont {Larson},
  \citenamefont {Hass}, \citenamefont {Ehrenreich},\ and\ \citenamefont
  {Carlsson}}]{Larson88}%
  \BibitemOpen
  \bibfield  {author} {\bibinfo {author} {\bibfnamefont {B.~E.}\ \bibnamefont
  {Larson}}, \bibinfo {author} {\bibfnamefont {K.~C.}\ \bibnamefont {Hass}},
  \bibinfo {author} {\bibfnamefont {H.}~\bibnamefont {Ehrenreich}}, \ and\
  \bibinfo {author} {\bibfnamefont {A.~E.}\ \bibnamefont {Carlsson}},\ }\href
  {\doibase 10.1103/PhysRevB.37.4137} {\bibfield  {journal} {\bibinfo
  {journal} {Phys. Rev. B}\ }\textbf {\bibinfo {volume} {37}},\ \bibinfo
  {pages} {4137} (\bibinfo {year} {1988})}\BibitemShut {NoStop}%
\bibitem [{\citenamefont {Larson}\ and\ \citenamefont
  {Ehrenreich}(1989)}]{Larson89}%
  \BibitemOpen
  \bibfield  {author} {\bibinfo {author} {\bibfnamefont {B.~E.}\ \bibnamefont
  {Larson}}\ and\ \bibinfo {author} {\bibfnamefont {H.}~\bibnamefont
  {Ehrenreich}},\ }\href {\doibase 10.1103/PhysRevB.39.1747} {\bibfield
  {journal} {\bibinfo  {journal} {Phys. Rev. B}\ }\textbf {\bibinfo {volume}
  {39}},\ \bibinfo {pages} {1747} (\bibinfo {year} {1989})}\BibitemShut
  {NoStop}%
\bibitem [{\citenamefont {Kuzian}\ \emph {et~al.}(2010)\citenamefont {Kuzian},
  \citenamefont {Laguta}, \citenamefont {Daré}, \citenamefont {Kondakova},
  \citenamefont {Marysko}, \citenamefont {Raymond}, \citenamefont {Garmash},
  \citenamefont {Pavlikov}, \citenamefont {Tkach}, \citenamefont {Vilarinho},\
  and\ \citenamefont {Hayn}}]{Kuzian10}%
  \BibitemOpen
  \bibfield  {author} {\bibinfo {author} {\bibfnamefont {R.~O.}\ \bibnamefont
  {Kuzian}}, \bibinfo {author} {\bibfnamefont {V.~V.}\ \bibnamefont {Laguta}},
  \bibinfo {author} {\bibfnamefont {A.-M.}\ \bibnamefont {Daré}}, \bibinfo
  {author} {\bibfnamefont {I.~V.}\ \bibnamefont {Kondakova}}, \bibinfo {author}
  {\bibfnamefont {M.}~\bibnamefont {Marysko}}, \bibinfo {author} {\bibfnamefont
  {L.}~\bibnamefont {Raymond}}, \bibinfo {author} {\bibfnamefont {E.~P.}\
  \bibnamefont {Garmash}}, \bibinfo {author} {\bibfnamefont {V.~N.}\
  \bibnamefont {Pavlikov}}, \bibinfo {author} {\bibfnamefont {A.}~\bibnamefont
  {Tkach}}, \bibinfo {author} {\bibfnamefont {P.~M.}\ \bibnamefont
  {Vilarinho}}, \ and\ \bibinfo {author} {\bibfnamefont {R.}~\bibnamefont
  {Hayn}},\ }\href {http://stacks.iop.org/0295-5075/92/i=1/a=17007} {\bibfield
  {journal} {\bibinfo  {journal} {EPL (Europhysics Letters)}\ }\textbf
  {\bibinfo {volume} {92}},\ \bibinfo {pages} {17007} (\bibinfo {year}
  {2010})}\BibitemShut {NoStop}%
\bibitem [{\citenamefont {Kuzian}\ \emph {et~al.}(2011)\citenamefont {Kuzian},
  \citenamefont {Dar\'e}, \citenamefont {Savoyant}, \citenamefont
  {D'Ambrosio},\ and\ \citenamefont {Stepanov}}]{Kuzian11}%
  \BibitemOpen
  \bibfield  {author} {\bibinfo {author} {\bibfnamefont {R.~O.}\ \bibnamefont
  {Kuzian}}, \bibinfo {author} {\bibfnamefont {A.~M.}\ \bibnamefont {Dar\'e}},
  \bibinfo {author} {\bibfnamefont {A.}~\bibnamefont {Savoyant}}, \bibinfo
  {author} {\bibfnamefont {S.}~\bibnamefont {D'Ambrosio}}, \ and\ \bibinfo
  {author} {\bibfnamefont {A.}~\bibnamefont {Stepanov}},\ }\href {\doibase
  10.1103/PhysRevB.84.165207} {\bibfield  {journal} {\bibinfo  {journal} {Phys.
  Rev. B}\ }\textbf {\bibinfo {volume} {84}},\ \bibinfo {pages} {165207}
  (\bibinfo {year} {2011})}\BibitemShut {NoStop}%
\bibitem [{\citenamefont {Slater}\ and\ \citenamefont {Koster}(1954)}]{SK}%
  \BibitemOpen
  \bibfield  {author} {\bibinfo {author} {\bibfnamefont {J.~C.}\ \bibnamefont
  {Slater}}\ and\ \bibinfo {author} {\bibfnamefont {G.~F.}\ \bibnamefont
  {Koster}},\ }\href {\doibase 10.1103/PhysRev.94.1498} {\bibfield  {journal}
  {\bibinfo  {journal} {Phys. Rev.}\ }\textbf {\bibinfo {volume} {94}},\
  \bibinfo {pages} {1498} (\bibinfo {year} {1954})}\BibitemShut {NoStop}%
\bibitem [{\citenamefont {Boekema}\ \emph {et~al.}(1972)\citenamefont
  {Boekema}, \citenamefont {Van~der Waude},\ and\ \citenamefont
  {Sawatzky}}]{Boekema72}%
  \BibitemOpen
  \bibfield  {author} {\bibinfo {author} {\bibfnamefont {C.}~\bibnamefont
  {Boekema}}, \bibinfo {author} {\bibfnamefont {F.}~\bibnamefont {Van~der
  Waude}}, \ and\ \bibinfo {author} {\bibfnamefont {G.~A.}\ \bibnamefont
  {Sawatzky}},\ }\href@noop {} {\bibfield  {journal} {\bibinfo  {journal} {Int.
  J. Magnetism}\ }\textbf {\bibinfo {volume} {3}},\ \bibinfo {pages} {341}
  (\bibinfo {year} {1972})}\BibitemShut {NoStop}%
\bibitem [{\citenamefont {Harrison}(1980)}]{Harrison}%
  \BibitemOpen
  \bibfield  {author} {\bibinfo {author} {\bibfnamefont {W.~A.}\ \bibnamefont
  {Harrison}},\ }\href@noop {} {\emph {\bibinfo {title} {Electronic Structure
  and the Properties of Solids}}}\ (\bibinfo  {publisher} {Freeman},\ \bibinfo
  {address} {San Francisco},\ \bibinfo {year} {1980})\BibitemShut {NoStop}%
\bibitem [{\citenamefont {Schmidt}\ \emph {et~al.}(2011)\citenamefont
  {Schmidt}, \citenamefont {Lohmann},\ and\ \citenamefont
  {Richter}}]{Schmidt11}%
  \BibitemOpen
  \bibfield  {author} {\bibinfo {author} {\bibfnamefont {H.-J.}\ \bibnamefont
  {Schmidt}}, \bibinfo {author} {\bibfnamefont {A.}~\bibnamefont {Lohmann}}, \
  and\ \bibinfo {author} {\bibfnamefont {J.}~\bibnamefont {Richter}},\ }\href
  {\doibase 10.1103/PhysRevB.84.104443} {\bibfield  {journal} {\bibinfo
  {journal} {Phys. Rev. B}\ }\textbf {\bibinfo {volume} {84}},\ \bibinfo
  {pages} {104443} (\bibinfo {year} {2011})}\BibitemShut {NoStop}%
\bibitem [{hte()}]{hte}%
  \BibitemOpen
  \href {http://www.uni-magdeburg.de/jschulen/HTE/} {}\bibinfo {note} {We have
  used the 2011-09-23 version of HTE package available at
  http://www.uni-magdeburg.de/jschulen/HTE/}\BibitemShut {NoStop}%
\bibitem [{\citenamefont {Geller}\ and\ \citenamefont
  {Gilleo}(1957)}]{Geller57}%
  \BibitemOpen
  \bibfield  {author} {\bibinfo {author} {\bibfnamefont {S.}~\bibnamefont
  {Geller}}\ and\ \bibinfo {author} {\bibfnamefont {M.}~\bibnamefont
  {Gilleo}},\ }\href {\doibase http://dx.doi.org/10.1016/0022-3697(57)90044-6}
  {\bibfield  {journal} {\bibinfo  {journal} {Journal of Physics and Chemistry
  of Solids}\ }\textbf {\bibinfo {volume} {3}},\ \bibinfo {pages} {30 }
  (\bibinfo {year} {1957})}\BibitemShut {NoStop}%
\bibitem [{\citenamefont {Cherepanov}\ \emph {et~al.}(1993)\citenamefont
  {Cherepanov}, \citenamefont {Kolokolov},\ and\ \citenamefont
  {L'vov}}]{Cherepanov93}%
  \BibitemOpen
  \bibfield  {author} {\bibinfo {author} {\bibfnamefont {V.}~\bibnamefont
  {Cherepanov}}, \bibinfo {author} {\bibfnamefont {I.}~\bibnamefont
  {Kolokolov}}, \ and\ \bibinfo {author} {\bibfnamefont {V.}~\bibnamefont
  {L'vov}},\ }\href {\doibase http://dx.doi.org/10.1016/0370-1573(93)90107-O}
  {\bibfield  {journal} {\bibinfo  {journal} {Physics Reports}\ }\textbf
  {\bibinfo {volume} {229}},\ \bibinfo {pages} {81 } (\bibinfo {year}
  {1993})}\BibitemShut {NoStop}%
\bibitem [{\citenamefont {Lieb}\ and\ \citenamefont {Mattis}(1962)}]{Lieb62}%
  \BibitemOpen
  \bibfield  {author} {\bibinfo {author} {\bibfnamefont {E.}~\bibnamefont
  {Lieb}}\ and\ \bibinfo {author} {\bibfnamefont {D.}~\bibnamefont {Mattis}},\
  }\href {\doibase 10.1063/1.1724276} {\bibfield  {journal} {\bibinfo
  {journal} {Journal of Mathematical Physics}\ }\textbf {\bibinfo {volume}
  {3}},\ \bibinfo {pages} {749} (\bibinfo {year} {1962})}\BibitemShut {NoStop}%
\bibitem [{\citenamefont {Raevskii}\ \emph {et~al.}(2002)\citenamefont
  {Raevskii}, \citenamefont {Sarychev}, \citenamefont {Bryugeman},
  \citenamefont {Reznichenko}, \citenamefont {Shilkina}, \citenamefont
  {Razumovskaya}, \citenamefont {Nikolaev}, \citenamefont {Vyshatko},\ and\
  \citenamefont {Salak}}]{Raevskii02}%
  \BibitemOpen
  \bibfield  {author} {\bibinfo {author} {\bibfnamefont {I.}~\bibnamefont
  {Raevskii}}, \bibinfo {author} {\bibfnamefont {D.}~\bibnamefont {Sarychev}},
  \bibinfo {author} {\bibfnamefont {S.}~\bibnamefont {Bryugeman}}, \bibinfo
  {author} {\bibfnamefont {L.}~\bibnamefont {Reznichenko}}, \bibinfo {author}
  {\bibfnamefont {L.}~\bibnamefont {Shilkina}}, \bibinfo {author}
  {\bibfnamefont {O.}~\bibnamefont {Razumovskaya}}, \bibinfo {author}
  {\bibfnamefont {V.}~\bibnamefont {Nikolaev}}, \bibinfo {author}
  {\bibfnamefont {N.}~\bibnamefont {Vyshatko}}, \ and\ \bibinfo {author}
  {\bibfnamefont {A.}~\bibnamefont {Salak}},\ }\href {\doibase
  10.1134/1.1523519} {\bibfield  {journal} {\bibinfo  {journal}
  {Crystallography Reports}\ }\textbf {\bibinfo {volume} {47}},\ \bibinfo
  {pages} {1012} (\bibinfo {year} {2002})}\BibitemShut {NoStop}%
\bibitem [{\citenamefont {Pirnie}\ \emph {et~al.}(1966)\citenamefont {Pirnie},
  \citenamefont {Wood},\ and\ \citenamefont {Eve}}]{Pirnie66}%
  \BibitemOpen
  \bibfield  {author} {\bibinfo {author} {\bibfnamefont {K.}~\bibnamefont
  {Pirnie}}, \bibinfo {author} {\bibfnamefont {P.}~\bibnamefont {Wood}}, \ and\
  \bibinfo {author} {\bibfnamefont {J.}~\bibnamefont {Eve}},\ }\href@noop {}
  {\bibfield  {journal} {\bibinfo  {journal} {Mol. Phys.}\ }\textbf {\bibinfo
  {volume} {11}},\ \bibinfo {pages} {551} (\bibinfo {year} {1966})}\BibitemShut
  {NoStop}%
\bibitem [{\citenamefont {Juh\'asz~Junger}\ \emph {et~al.}(2009)\citenamefont
  {Juh\'asz~Junger}, \citenamefont {Ihle},\ and\ \citenamefont
  {Richter}}]{Junger09}%
  \BibitemOpen
  \bibfield  {author} {\bibinfo {author} {\bibfnamefont {I.}~\bibnamefont
  {Juh\'asz~Junger}}, \bibinfo {author} {\bibfnamefont {D.}~\bibnamefont
  {Ihle}}, \ and\ \bibinfo {author} {\bibfnamefont {J.}~\bibnamefont
  {Richter}},\ }\href {\doibase 10.1103/PhysRevB.80.064425} {\bibfield
  {journal} {\bibinfo  {journal} {Phys. Rev. B}\ }\textbf {\bibinfo {volume}
  {80}},\ \bibinfo {pages} {064425} (\bibinfo {year} {2009})}\BibitemShut
  {NoStop}%
\bibitem [{\citenamefont {Yasuda}\ \emph {et~al.}(2005)\citenamefont {Yasuda},
  \citenamefont {Todo}, \citenamefont {Hukushima}, \citenamefont {Alet},
  \citenamefont {Keller}, \citenamefont {Troyer},\ and\ \citenamefont
  {Takayama}}]{Yasuda05}%
  \BibitemOpen
  \bibfield  {author} {\bibinfo {author} {\bibfnamefont {C.}~\bibnamefont
  {Yasuda}}, \bibinfo {author} {\bibfnamefont {S.}~\bibnamefont {Todo}},
  \bibinfo {author} {\bibfnamefont {K.}~\bibnamefont {Hukushima}}, \bibinfo
  {author} {\bibfnamefont {F.}~\bibnamefont {Alet}}, \bibinfo {author}
  {\bibfnamefont {M.}~\bibnamefont {Keller}}, \bibinfo {author} {\bibfnamefont
  {M.}~\bibnamefont {Troyer}}, \ and\ \bibinfo {author} {\bibfnamefont
  {H.}~\bibnamefont {Takayama}},\ }\href {\doibase
  10.1103/PhysRevLett.94.217201} {\bibfield  {journal} {\bibinfo  {journal}
  {Phys. Rev. Lett.}\ }\textbf {\bibinfo {volume} {94}},\ \bibinfo {pages}
  {217201} (\bibinfo {year} {2005})}\BibitemShut {NoStop}%
\bibitem [{\citenamefont {Pietrzak}\ \emph {et~al.}(1981)\citenamefont
  {Pietrzak}, \citenamefont {Maryanowska},\ and\ \citenamefont
  {Leciejewicz}}]{Pietrzak81}%
  \BibitemOpen
  \bibfield  {author} {\bibinfo {author} {\bibfnamefont {J.}~\bibnamefont
  {Pietrzak}}, \bibinfo {author} {\bibfnamefont {A.}~\bibnamefont
  {Maryanowska}}, \ and\ \bibinfo {author} {\bibfnamefont {J.}~\bibnamefont
  {Leciejewicz}},\ }\href {\doibase 10.1002/pssa.2210650164} {\bibfield
  {journal} {\bibinfo  {journal} {physica status solidi (a)}\ }\textbf
  {\bibinfo {volume} {65}},\ \bibinfo {pages} {K79} (\bibinfo {year}
  {1981})}\BibitemShut {NoStop}%
\bibitem [{\citenamefont {Ivanov}\ \emph {et~al.}(2000)\citenamefont {Ivanov},
  \citenamefont {Tellgren}, \citenamefont {Rundlof}, \citenamefont {Thomas},\
  and\ \citenamefont {Ananta}}]{Ivanov00}%
  \BibitemOpen
  \bibfield  {author} {\bibinfo {author} {\bibfnamefont {S.~A.}\ \bibnamefont
  {Ivanov}}, \bibinfo {author} {\bibfnamefont {R.}~\bibnamefont {Tellgren}},
  \bibinfo {author} {\bibfnamefont {H.}~\bibnamefont {Rundlof}}, \bibinfo
  {author} {\bibfnamefont {N.~W.}\ \bibnamefont {Thomas}}, \ and\ \bibinfo
  {author} {\bibfnamefont {S.}~\bibnamefont {Ananta}},\ }\href
  {http://stacks.iop.org/0953-8984/12/i=11/a=305} {\bibfield  {journal}
  {\bibinfo  {journal} {Journal of Physics: Condensed Matter}\ }\textbf
  {\bibinfo {volume} {12}},\ \bibinfo {pages} {2393} (\bibinfo {year}
  {2000})}\BibitemShut {NoStop}%
\bibitem [{\citenamefont {Morrish}(1980)}]{Morrish}%
  \BibitemOpen
  \bibfield  {author} {\bibinfo {author} {\bibfnamefont {A.~H.}\ \bibnamefont
  {Morrish}},\ }\href@noop {} {\emph {\bibinfo {title} {Physical principles of
  magnetism}}}\ (\bibinfo  {publisher} {R.E. Krieger Publishing Company,
  Inc.},\ \bibinfo {address} {Huntington},\ \bibinfo {year} {1980})\BibitemShut
  {NoStop}%
\end{thebibliography}%
\end{document}